\documentclass[iop,apj,numberedappendix]{emulateapj}

\usepackage{graphicx}
\usepackage{natbib}
\usepackage{amssymb}
\usepackage{amsmath}
\usepackage{hyperref}

\newcommand{\RA}{{\it{}RadioAstron}\ }
\newcommand{\psr}{PSR~B0329+54\ }

\newcommand{\convolve}{\operatornamewithlimits{\otimes}}

\makeindex \citeindextrue

\received{January 18, 2015}
\accepted{March 15, 2016}
\journalinfo{Astrophysical~journal}
\submitted{}

\shorttitle{RadioAstron discovery of scaterring substructure in PSR~0329+54}
\shortauthors{Gwinn et al.}

\begin{document}

\title{PSR B0329+54: Statistics of Substructure Discovered within the Scattering Disk on RadioAstron Baselines of up to 235,000 km}

\author{
C.~R.\ Gwinn\altaffilmark{1},
M.~V.\ Popov\altaffilmark{2},
N.\ Bartel\altaffilmark{3},
A.~S.\ Andrianov\altaffilmark{2},
M.~D.\ Johnson\altaffilmark{4},
B.~C.\ Joshi\altaffilmark{5},
N.~S.\ Kardashev\altaffilmark{2},
\mbox{R.\ Karuppusamy\altaffilmark{6},}
Y.~Y.\ Kovalev\altaffilmark{1,6},
M.\ Kramer\altaffilmark{6},
A.~G.\ Rudnitskii\altaffilmark{2},
E.~R.\ Safutdinov\altaffilmark{2},
V.~I.\ Shishov\altaffilmark{7},
\mbox{T.~V.\ Smirnova\altaffilmark{7},}
V.~A.\ Soglasnov\altaffilmark{2},
S.\ F.\ Steinmassl\altaffilmark{8},
J.~A.\ Zensus\altaffilmark{6},
V.~I.\ Zhuravlev\altaffilmark{2}
}

\altaffiltext{1}{University of California at Santa Barbara, Santa Barbara, CA 93106-4030, USA} 
\altaffiltext{2}{Astro Space Center of Lebedev Physical Institute, Profsoyuznaya 84/32, Moscow 117997, Russia}
\altaffiltext{3}{York University, 4700 Keele St., Toronto, ON M3J 1P3, Canada}
\altaffiltext{4}{Harvard-Smithsonian Center for Astrophysics, 60 Garden St, Cambridge, MA 02138, USA}
\altaffiltext{5}{National Centre for Radio Astrophysics, Post Bag 3, Ganeshkhind, Pune 411007, India}
\altaffiltext{6}{Max-Planck-Institut f\"ur Radioastronomie, Auf dem H\"ugel 69, Bonn 53121, Germany}
\altaffiltext{7}{Pushchino Radio Astronomy Observatory, Astro Space Center of Lebedev Physical Institute, Pushchino 142290, Moscow region, Russia}
\altaffiltext{8}{Physik-Department, Technische Universit\"at M\"unchen, James Franck-Strasse 1, Garching bei M\"unchen 85748, Germany}

\begin{abstract}
We discovered fine-scale structure within the scattering disk of PSR~B0329+54 in observations with the \RA ground-space radio interferometer.
Here, we describe this phenomenon, characterize it with averages and correlation functions, and interpret it as the result of decorrelation of the impulse-response function of interstellar scattering between the widely-separated antennas.
This instrument included
the 10-m Space Radio Telescope, the 110-m Green Bank Telescope, the $14\times25$-m Westerbork Synthesis Radio Telescope, and the 64-m Kalyazin Radio Telescope.
The observations were performed at 324 MHz, on baselines of up to 235,000~km in November~2012 and January~2014. 
In the delay domain, on long baselines the interferometric visibility consists of many discrete spikes within a limited range of delays.
On short baselines it consists of a sharp spike surrounded by lower spikes.
The average envelope of correlations of the visibility function show two exponential scales, with characteristic delays of 
$\tau_1=4.1\pm 0.3\ \mu{\rm s}$ and $\tau_2=23\pm 3\ \mu{\rm s}$, indicating the presence of two scales of scattering in the interstellar medium. 
These two scales are present in the pulse-broadening function.
The longer scale contains 0.38 times the scattered power of the shorter one.
We suggest that the longer tail arises from highly-scattered paths, possibly from anisotropic scattering or from substructure at large angles.
\end{abstract}


\section{Introduction}
\label{intro}

All radio signals from cosmic sources are distorted by the plasma
turbulence in the interstellar medium (ISM). Understanding of this
turbulence is therefore essential for the proper interpretation of
astronomical radio observations. The properties and characteristics
of this turbulence can best be studied by observing point-like
radio sources, where the results are not influenced by the extended
structure of the source, but instead are directly attributable to the
effect of the ISM itself. Pulsars are such sources. 
Dispersion and scattering
affect radio emission from pulsars.
Whereas dispersion in the plasma column introduces delays in arrival time
that depend upon frequency and results in smearing of the pulse, scattering by density inhomogeneities causes angular broadening,
pulse broadening, intensity modulation or scintillation, and
distortion of radio spectra in the form of diffraction patterns.
The scattering effects have already been studied extensively theoretically \citep[see, e.g.,][]{prokhorov1975,rickett1977,goodman1989, narayan1989,
shishov2003} and observationally with ground VLBI of Sgr\,A$^*$ \citep{gwinn2014} and pulsars \citep[see, e.g.,][]{bartel1985,desai1992,kondratiev2007}, 
as well as with ground-space VLBI of \psr \citep[\textit{Halca},][]{yangalov2001} and the quasar 3C\,273 \citep[\textit{RadioAstron},][]{3C273_RA}.
Whereas the VSOP pulsar observations were done at a relatively high
frequency of 1.7~GHz and with baselines of $\approx$25,000~km and less,
ground-space VLBI with \RA allows observations at one-fifth the frequency, where propagation effects are expected to be
much stronger, and with baselines ${\sim}10$ times longer \citep{kardashev2013}. Such observations can resolve the scatter-broadened image of a pulsar and reveal new information about the scattering medium \citep{smirnova2014}.

In this paper, we study the scattered image of the pulsar B0329+54 with {\it RadioAstron}.
We demonstrate that the pulsar is detected on baselines that fully resolve the scattering disk.
The interferometric visibility on these long baselines takes the form of random phase and amplitude variations that vary randomly with observing frequency and time.
In the Fourier-conjugate domain of delay and fringe rate, the visibility forms a localized, extended region around the origin, composed of many random spikes.
We characterize the shape of this region using averages and correlation functions.
We argue theoretically that its extent in delay is given by the average envelope of the impulse-response function of interstellar scattering,
sometimes called the pulse-broadening function.
We find that the observed distribution is well-fit by a model that is derived from an impulse-response function that has two different exponential scales.
We discuss possible origins of the two scales.

\begin{deluxetable*}{llllr}
\tabletypesize{\small}
\tablecaption{Diary of observations}
\tablewidth{0pt}
\tablehead{
        \colhead{Epoch of}         & \colhead{Time}& \colhead{Ground}          & \colhead{Polarizations} &\colhead{Scan}\\
         \colhead{Observations} & \colhead{Span}& \colhead{ Telescopes}   &                                      & \colhead{Length}
}
\startdata
2012 Nov 26 through 29   & 1\ hr/day                         & GB                              & RCP$+$LCP                &570\ s \\                
2014 Jan 1 and 2              &12 hr                                & WB, KL            & RCP                             &1170\ s  
\enddata
\label{tab:diary}
\end{deluxetable*}

\section{Theoretical Background}\label{sec:Theory_Intro}

Our fundamental observable is the interferometric visibility $V$.
In the domain of frequency $\nu$, this is the product of electric fields at two antennas $A$ and $B$:
\begin{align}
\tilde V_{AB} (\nu ,t) =\tilde E_A(\nu ,t) \tilde E^*_B (\nu , t) .
\label{eq:Vdef}
\end{align}
This representation of the visibility is known as the cross spectrum, or cross-power spectrum.  Because electric fields at the antennas are complex
and different, $\tilde V_{AB}$ is complex. 
Usually visibility is averaged over multiple accumulations of the spectrum, to reduce noise from background and the noiselike electric field of the source.
The second argument $t$ allows for the possibility that the visibility changes in time, as it does for a scintillating source,
over times longer than the time to accumulate a single spectrum.
Such a spectrum that changes in time is known as a ``dynamic spectrum'' \citep{bracewell2000}.
The correlator used to analyze our data, as discussed in Sections\ \ref{sec:obs} and\ \ref{sec:red}, calculates $\tilde V_{AB} (\nu ,t)$ \citep{RACorrelator}.
Hereafter, we omit the baseline subscript indicating baseline $AB$ in this paper, except in sections of the Appendix where the baseline is important.

\begin{figure}
\centering
\includegraphics[width=0.48\textwidth]{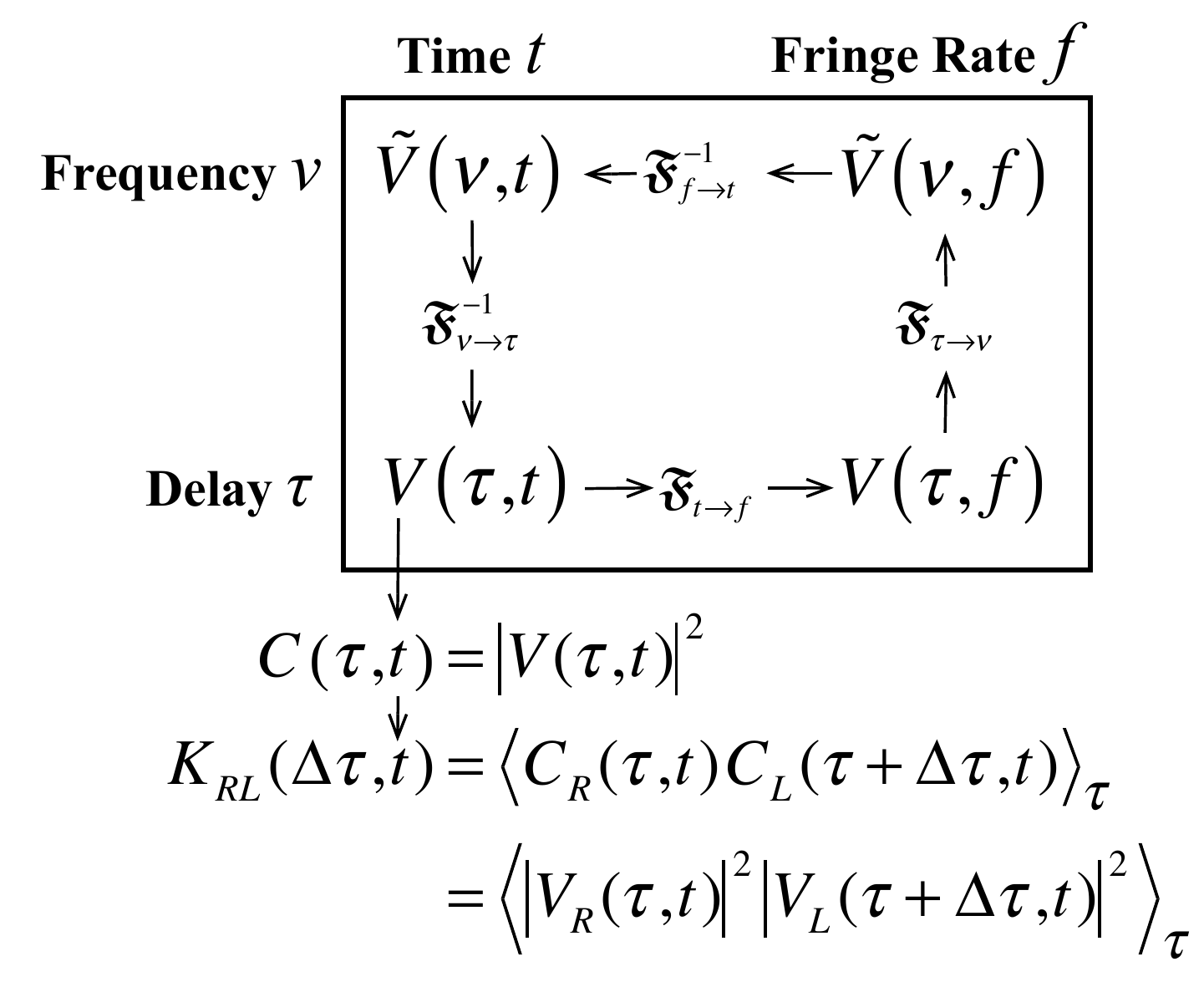}
\caption{Relations among the interferometric visibility $V$ in various domains,
and functions derived from it.
The fundamental observable is the visibility in the domain of frequency $\nu$ and time $t$, $\tilde V(\nu,t)$;
this is known as the cross-power spectrum, or cross spectrum.
An inverse Fourier transform of $\nu$ to delay $\tau$ leads to the visibility $V(\tau, t)$;
this is the cross-correlation function of electric fields in the time domain (see Equation\ \ref{eq:Vtau_def}).
A forward Fourier transform of $t$ to fringe rate $f$ leads to $V(\tau, f)$.
A forward transform of $\tau$ back to $\nu$ produces $\tilde V(\nu,f)$,
and an inverse Fourier transform of $f$ to $t$ returns to $\tilde V(\nu,t)$
The square modulus of $V(\tau, t)$ is $C(\tau,t)$.
The cross-correlation function in $\tau$ of $C_R$ for right- and $C_L$ for left-circular polarization is $K_{RL}(\Delta \tau, t)$.
We denote the Fourier transform by $\mathfrak F$, and quantities in the domain of frequency $\nu$ by the accent $\tilde\ $.
\label{fig:FTCircle}}
\end{figure}

Under the assumptions that the source is pointlike, and that we can ignore background and source noise,
the impulse-response function of interstellar scattering $g$ determines the visibility of the source.
A single delta-function impulse of electric field at the source is received as a function $g(t_e)$ of time $t_e$ at the observer.
Here, $t_e$ is Fourier-conjugate to $\nu$ and varies at the Nyquist rate.
The visibility is the product of Fourier transforms of $g$ at the two antennas: 
\begin{align}
\tilde V_{AB}  = \tilde g_A \tilde g_B^*
\end{align}
where $\tilde g$ is the Fourier transform of $g(t_e)$.

We denote the typical duration of $g(t_e)$ as $\tau_{\rm sc}$, the broadening time for a sharp pulse.
Within this time span, $g(t_e)$ has a complicated amplitude and phase.
The function $g(t_e)$ changes over longer times,
as the line of sight shifts with motions of source, observer, and medium.
This change takes place on a timescale $t_{\rm sc}$,
and over a spatial scale $S_{\rm sc}$.
The shorter and longer timescales $\tau_{\rm sc}$ and $t_{sc}$ lead to our use of dual time variables: $t_e$, of up to a few times $\tau_{sc}$ and Fourier-conjugate to $\nu$; and $t$, of a fraction of $t_{sc}$ or more and Fourier-conjugate to $f$.
This duality is commonly expressed via the ``dynamic spectrum'' (see Section\ \ref{sec:AppImpluseResponseFunction}).
If the scattering material remains nearly at rest while the line of sight travels through it at velocity $V_{\perp}$,
then one spatial dimension in the observer plane maps into time, and 
\begin{align}
t_{\rm sc} &= S_{\rm sc}/V_{\perp}
\label{eq:t_sc}
\end{align}
The averaged square modulus of $g$ is the pulse-broadening function $G=\langle g(t_e) g(t_e)^*\rangle_S$.
Here, the subscripted angular brackets $\langle...\rangle_S$ indicate an average over realizations of the scattering.
This function is the average observed intensity for a single sharp pulse emitted at the source.
An average over time is usually assumed to approximate the desired average over an ensemble of statistically-identical realizations of scattering. 

We derive a number of 
representations of the visibility and quantities derived from it, and show that these provide straightforward means to extract the impulse-response function.
These functions are summarized in Figure~\ref{fig:FTCircle},
and discussed briefly here, and in detail in Section~\ref{sec:impulse-response_visibility} of the Appendix.
In particular, visibility in the domain of delay $\tau$ and time $t$ is $V(\tau, t)$.
This is the correlation function of electric field at the two antennas $A$ and $B$ (Equation\ \ref{eq:Vtau_def}),
and is the inverse Fourier transform of $\tilde V(\nu, t)$ from $\nu$ to $\tau$.
We are also concerned with the square modulus of $V(\tau, t)$ (see\ Section\ \ref{sec:acf_visibility}):
\begin{align}
C (\tau, t) &= \left| V(\tau, t) \right|^2
\end{align}
We calculate $C$ for right- and left-circular polarizations separately, and then correlate them in delay $\tau$ to form $K_{RL}$, the cross-correlation between polarizations:
\begin{align}
K_{RL} (\Delta\tau, t ) &=\textstyle{\frac{1}{N}} \sum_{\tau} C_R (\tau, t) C_L (\tau+\Delta \tau, t)  
\label{eq:KRLdef}
\end{align}
Here, $K_{RL}$ is the correlation of a single measurement of $C_R$ and $C_L$, and $N$ is the number of samples in $C_R$ and $C_L$.

When averaged over many realizations of the scattering material,
$\langle K_{RL}\rangle_S$ is related to the statistics of the pulse-broadening function $G$.
Most commonly, the average over many realizations of scattering material is approximated by averaging over a time much longer than $t_{\rm sc}$;
for this reason we omit the time argument for $\langle K_{RL}(\Delta\tau)\rangle_S$.
Equivalently, evaluation of $\langle K_{RL}(\Delta \tau,f_{max})\rangle = \langle C_R (\tau, f_{max}) C_L (\tau+\Delta \tau, f_{max})\rangle_{\tau}$ at the fringe rate $f_{max}$ of the maximum magnitude of $K_{RL}$
yields the same time average.
For this theoretical discussion, $f_{max}=0$; for practical observations, instrumental factors can offset the fringe rate from zero, so that $f_{max}$ provides the most reliable time average.

For a baseline that extends much further than the scale of scattering $S_{\rm sc}$
(see Equation\ \ref{eq:K_result}):
\begin{align}
\langle K_{RL} (\tau ) \rangle_S &=
G  (\tau )\otimes G_-  (\tau )\otimes G  (\tau )\otimes G_-  (\tau )\\
&\quad +
\big(1 {\rm\ if\ }\tau=0
\big) \nonumber
\end{align}
Here, we introduce the symbol $\otimes$ to indicate convolution, and denote the time-reverse of $G$ as $G_-(\tau) = G(-\tau)$.

Our analysis method differs somewhat from \citet{smirnova2014}, who used structure functions of intensity, visibility, and visibility squared
to study scattering of pulsar B0950+08 on an extremely long baseline to {\it RadioAstron}.  
The two methods are closely related theoretically. Structure functions are particularly valuable when the 
characteristic bandwidth approaches the instrumental bandwidth, and can be extended to cases where the signal-to-noise ratio is low,
as they discuss.

\begin{deluxetable}{lcc}
\tabletypesize{\small}
\tablecaption{Observations on Earth-Space Baselines}
\tablewidth{0pt}
\tablehead{
\colhead{Epoch}           & \colhead{Projected}          & \colhead{RA}\\
                       & \colhead{Baseline Length}             & \colhead{Observing Time}\\
                       & \colhead{($10^3$\ km)}       & \colhead{(minutes)} 
}
\startdata
2012 Nov 26            &\phantom{10}   60                         &\phantom{.} 60  \\                
2012 Nov 27            &\phantom{10}   90                         &\phantom{.} 60  \\                
2012 Nov 28            &\phantom{0} 175                          &\phantom{.} 60  \\                
2012 Nov 29            &\phantom{1} 235                          &\phantom{.} 60  \\                
2014 Jan 1              &\phantom{10} 20                            &\phantom{.} 60 \\
2014 Jan 2              &\phantom{10} 70                           & 100 \\
2014 Jan 2             &\phantom{10} 90                           & 120 
\enddata
\label{tab:baselines}
\end{deluxetable}

\section{Observations}
\label{sec:obs} 

The observations were made in two sessions: the first
for one hour each on the four successive days November 26 to 29,
2012, and the second for a total of 12 hours on the two days January 1
and 2, 2014. The first session used the 10-m \RA Space Radio
Telescope (RA) together with the 110-m Robert C.\ Byrd Green Bank
Telescope (GB). The second session used the RA together with the 
$14\times25$-m Westerbork Synthesis Radio Telescope (WB), and the 64-m Kalyazin
Radio Telescopes (KL). Both right (RCP) and left circular
polarizations (LCP) were recorded in November 2012, and only one
polarization channel (RCP) was recorded in January 2014. Because of
an RA peculiarity at 324~MHz, the
316--332~MHz observing band was recorded as a single upper sideband, with one-bit digitization
at the RA and with two-bit digitization at the GB, WB, and KL.
Science data from the RA were transmitted in real time to the
telemetry station in Pushchino \citep{kardashev2013} and then recorded with the
\RA data recorder (RDR). This type of recorder was also used
at the KL, while the Mk5B recording system was used at the GB and
WB. Table~\ref{tab:diary} summarizes the observations.

The data were transferred via internet to the Astro Space Center (ASC) in Moscow and then processed with the ASC correlator with gating and dedispersion applied \citep{RACorrelator}. To determine the phase of the gate in the pulsar period, the average pulse profile was computed for every station by integrating the autocorrelation spectra obtained from the ASC correlator. The autocorrelation spectra $V_{AA}(\nu,t)$ are the square modulus of electric field at a single antenna.

In November 2012  the projected baselines to the space radio telescope were about 60, 90, 175, and 235 thousand kilometers for the four consecutive days, respectively. Data were recorded in 570-second scans, with 30-second gaps between scans. In January 2014 the projected baselines were about 20, 70, and 90 thousand kilometers during the 12-hour session. Data were recorded in 1170-second scans. The RA operated only during three sets of scans of 60, 100 and 120 min each, with large gaps in between caused by  thermal constraints on the spacecraft. The auto-level (AGC), phase cal, and noise diode were turned off during our observations to avoid interference with pulses from the pulsar.  Table~\ref{tab:baselines} gives parameters of the 
Earth-space baselines observed.

\section{Data Reduction}
\label{sec:red}

\subsection{Correlation}\label{sec:Corr}

All of the recorded data were correlated with the ASC correlator using
4096 channels for the November 2012 session and 2048 channels for the January 2014
session, with gating and dedispersion activated. The ON-pulse window was centered on
the main component of the average profile, with a width of 5~ms in the November 2012 session and 8~ms in the January 2014 session. 
These compare with a 7-ms pulse width at 50\% of the peak flux density \citep{Lorimer1995}.
The
OFF-pulse window was offset from the main pulse by half a period and had the same width as the ON-pulse window.
The correlator output was always sampled synchronously with the pulsar
period of 0.714~s (single pulse mode). We used ephemerides
computed with the program TEMPO for the Earth center \citep{TEMPO}. The results of the
correlation were tabulated as cross power spectra,  $\tilde V(\nu,t)$, written in standard
FITS format.

\begin{deluxetable*}{rrrrrcc}
\tabletypesize{\small}
\tablecaption{Measured Scattering Parameters of \psr}
\tablewidth{0pt}
\tablehead{
\colhead{Epoch}  & \colhead{$t_{\rm sc}$}& \colhead{$\Delta\nu_{\rm sc}$}   &\colhead{$w_{n\tau}$}&\colhead{$w_{nf}$} &\colhead{$\tau_1=1/k_1$}  & \colhead{$\tau_2=1/k_2$} \\
                 & \colhead{(s)}           & \colhead{(kHz)}                &\colhead{(ns)}       &\colhead{(mHz)}    &\colhead{($\mu$s)} &\colhead{($\mu$s)} \\
\colhead{(1)}    & \colhead{(2)}           & \colhead{(3)}                  &\colhead{(4)}        &\colhead{(5)}        & \colhead{(6)}     & \colhead{(7)}       }
\startdata
Nov 2012         & $114\pm2$               & $15\pm 2$                      & $50\pm5$            &  $20\pm 2$           & $4.1\pm 0.3$      &$23\pm 3$ \\    
Jan 2014         & $102\pm2$                 & $7\pm2$                      & $43\pm3$            &  $25\pm 3$           & $7.5\pm0.3$            &  -- 
\enddata
\tablecomments{
Columns are as follows:
(1) Date of observations,
(2) Scintillation time from autocorrelation spectra as the half width at 1/e of maximum,
(3) Scintillation bandwidth from single-dish autocorrelation spectra as the half-width at half maximum (HWHM),
(4) HWHM of a sinc function fit to the central spike of the visibility distribution along the delay axis,
(5) HWHM of a sinc function fit to the central spike of the visibility distribution along the fringe rate axis,
(6) Scale of the narrow component of $|K_{RL}(\Delta\tau)|$ (Section\ \ref{sec:two_scales}),
(7) Scale of the broad component of $|K_{RL}(\Delta\tau)|$ (Section\ \ref{sec:two_scales}).
}
\label{tab:scatt_params}
\end{deluxetable*}

\subsection{Single-Dish Data Reduction}\label{sec:SDproc}

Using autocorrelation spectra at GB, KL, and WB, we measured the scintillation time $t_{sc}$ and bandwidth $\Delta\nu_{sc} = 1/2\pi \tau_{sc}$.
The results are given in Table\ \ref{tab:scatt_params}.
Our analysis using interferometric data,
for which the noise baseline is absent and the spectral resolution was higher, is more accurate for the constants $\tau_1$ and $\tau_2$  as discussed below,
so we quote those values in Table\ \ref{tab:scatt_params}.

\begin{figure*}
\centering
\includegraphics[angle=0,trim=1cm 1cm 0cm 4.5cm,clip,width=\textwidth]{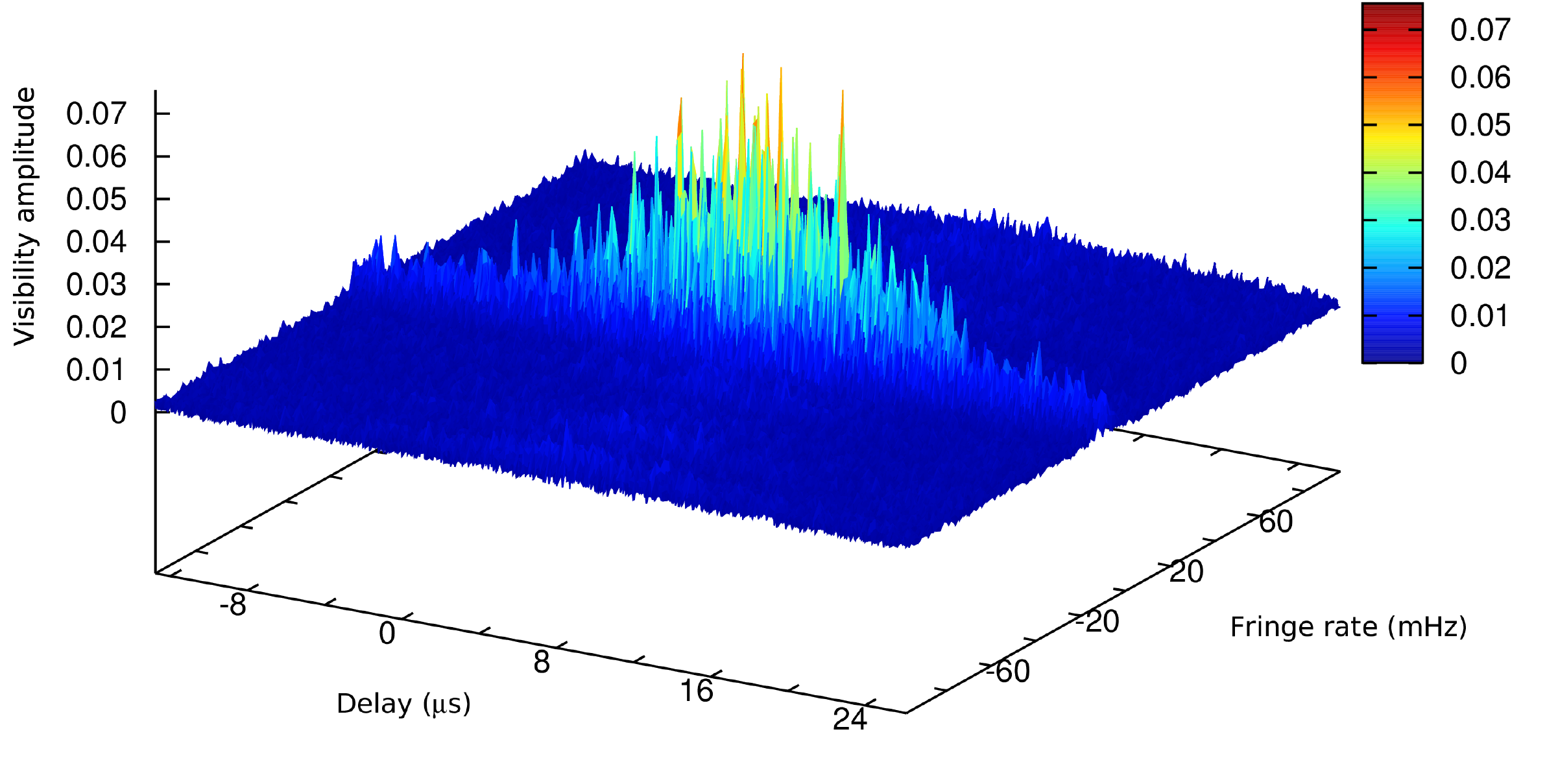}
\includegraphics[angle=0,trim=1cm 1cm 0cm 3cm,clip,width=\textwidth]{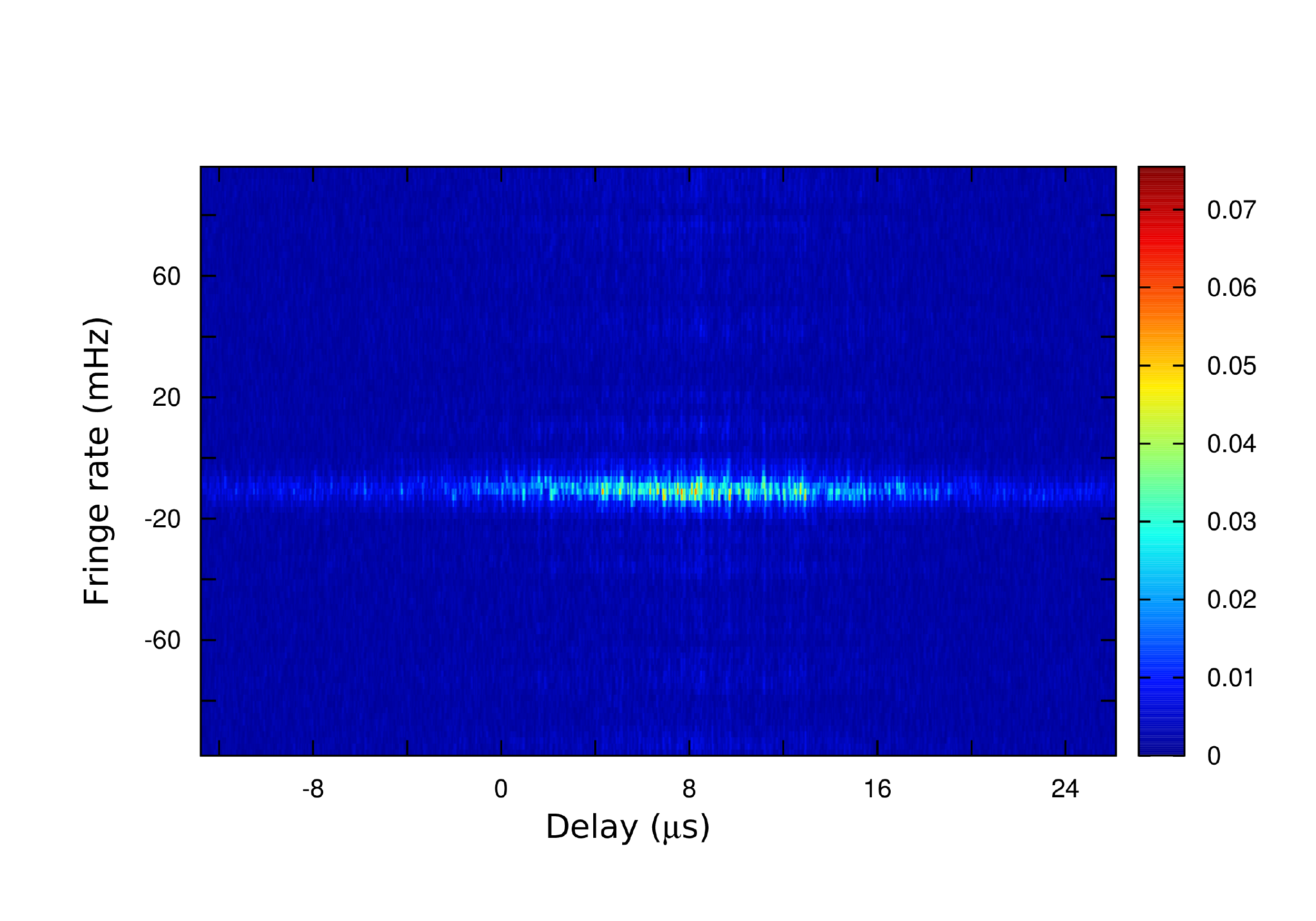}
\caption{Magnitude of visibility in the delay-fringe rate domain $|V(\tau,f)|$, for a 500-s
time span on 29 November 2012 in the RCP channel, on the RA-GB baseline. Visibility is normalized for autocorrelation: $|V(0,0)|=1$.
The axes show instrumental offsets, including about $6\ \mu{\rm s}$ in delay.
\textit{Top:} three-dimentional representation; \textit{bottom:} two-dimentional representation.}
\label{fig:frrt_delay}
\end{figure*}

\subsection{VLBI Data Reduction}\label{sec:VLBIproc}

The ASC correlator calculates the cross-power spectrum,
$\tilde V(\nu,t)$, as discussed in Sections\ \ref{sec:Theory_Intro} and\ \ref{sec:visibility}.
The resolution of the resulting cross-power spectra is 3.906\ kHz for the 2012 observations and 7.812\ kHz for the 2014
observations.  
Because the scintillation bandwidth was comparable to the channel bandwidth for the 2014 observations, as shown in Table\ \ref{tab:scatt_params},
and because the single recorded polarization at that epoch prevented us from correlating polarizations to form $K_{RL}$, as discussed in Section\ \ref{sec:two_scales},
we focus our analysis and interpretation on the 2012 observations.

\section{Analysis of Interferometric Visibility}\label{sec:IntVis}

We investigated the scattering of the pulsar from the visibility in the delay-fringe-rate domain, $V(\tau, f)$. 
We studied the statistics of visibility $V(\tau,f)$ as a function of delay, fringe rate, and baseline length. If there were no scattering material between the pulsar and the observer, we would expect for $|V(\tau,f)|$ one spike at zero delay and fringe rate with magnitude that remains constant as a function of baseline length, and with width 
equal to the inverse of the observed bandwidth in delay, and the inverse of the scan length in fringe rate. Scattering material in between changes this picture. First we expect the spike at zero delay and fringe rate to decrease in magnitude with increasing baseline length, perhaps to the point where it would become invisible. Second, we expect additional 
spikes to appear around the spike at zero delay and fringe rate. The distribution of these spikes give us invaluable information about the statistics of the scattering material. 

As we discuss in this section, 
we fitted models to the distribution of visibility,
as measured by the correlation function $K_{RL}$,
and thus derived scintillation parameters that describe the impulse-response function for propagation along the line of sight from the pulsar.
We also computed the maximum visibility as a function of projected baseline length, as we discuss in detail in a separate paper (Paper II: Popov et al., in preparation).

For strong single
pulses the visibility in the cross spectra, $\tilde V(\nu,t)$, had signal-to-noise ratios sufficiently large for a useful analysis. 
However, we decided to analyze the data from the time series of multiple pulses.
Fourier transform of the cross spectrum, $\tilde V(\nu,t)$, to the delay/fringe-rate domain yields $V(\nu,f)$ and concentrates the signal into a central region,
and thus provides a high signal-to-noise ratio. 
The sampling rate of individual cross spectra in the time series was the pulse period of 0.714\ s, as noted in Section\ \ref{sec:Corr}.
The time span of cross spectra used to form $V(\tau,f)$ varied, ranging from 71.4\ s to 570\ s, depending upon the application.


\subsection{Distribution of visibility}

In Figure~\ref{fig:frrt_delay} we display the magnitude of the visibility in the delay/fringe-rate domain, $|V(\tau,f)|$, for a 500-s time span. 
The data were obtained on 29 November 2012 in the RCP channel for a projected $200 {\rm M}\lambda$ GB-RA baseline.
The cross spectra, $\tilde V(\nu,t)$, from which we obtained $|V(\tau,f)|$ were sampled with 4096 spectral channels across the 16-MHz band, at the pulsar period of 0.714 s;
consequently, the resolution was 0.03125\ $\mu$s in delay, and 2\ mHz in fringe rate.
As Figure\ \ref{fig:frrt_delay} shows, no dominant central spike is visible at zero delay and fringe rate, as would be expected for an unresolved source. Our long baseline interferometer completely resolves the scattering disk. Instead we see a distribution of spikes around zero delay and fringe rate that is concentrated in a relatively limited region of the delay-fringe rate domain.
The locations of the various spikes appear to be random. 
Because the scattering disk is completely resolved on our long baseline, we conclude that 
the spikes are a consequence of random reinforcement or cancellation of paths to the different locations of the two telescopes, and hence interferometer phase.

In Figure~\ref{fig:frrt_delay}, the distribution of the magnitude of visibility is relatively broad along the delay axis and relatively narrow along the fringe rate axis.
The extent is limited in delay to about the inverse of the scintillation bandwidth, $\tau_{\rm sc}=1/2\pi \Delta \nu_{\rm sc}$;
and in fringe rate to about the inverse of the diffractive timescale $t_\mathrm{\rm diff}$.
Within this region, the visibility shows many narrow, discrete spikes.
If statistics of the random phase and amplitude of scintillation are Gaussian, 
and the phases of the Fourier transform randomize the different sums that comprise the visibility in the delay-fringe rate domain,
then the square modulus of $V(\tau,f)$ should be drawn from an exponential distribution, multiplied by the envelope defined by the deterministic 
part of the impulse-response function, as discussed in the Appendix.

\begin{figure}
\centering
\includegraphics[angle=0,trim=1cm 1cm 1.5cm 2cm,width=0.49\textwidth]{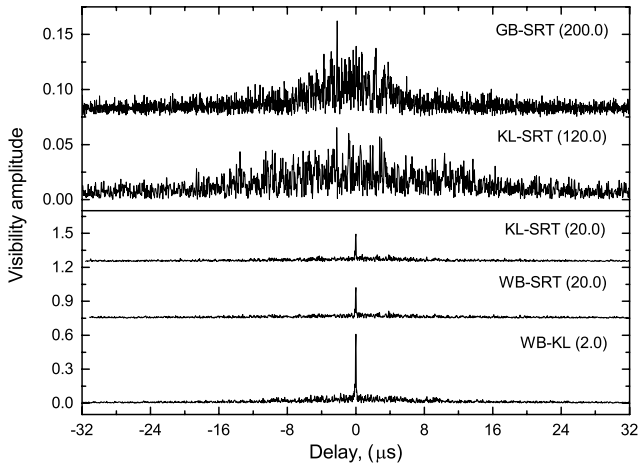}
\caption{Examples of the fine structure of the magnitude of visibility, $|V(\tau,f_{\rm max})|$, as a function of delay $\tau$, with fringe rate fixed at the maximum of the delay-fringe rate visibility near zero fringe rate, $f_{max}$ .
From lowermost to uppermost, the curves correspond to progressively longer baselines, 
with the telescopes indicated and the approximate baseline projections given in ${\rm M}\lambda$ in parentheses. 
Curves are offset vertically, and the upper 2 magnified as the vertical scale indicates, for ease of viewing.
All curves show 71.4\ s of data. Uppermost curve is from 2012 November 29;
the rest are from 2014 January, when multiple ground telescopes provided shorter baselines.
Note variation in scattering time between epochs as given in Table\ \ref{tab:scatt_params}.
The uppermost panel is the cross-section of the data shown in Figure~\ref{fig:frrt_delay},
but for 71.4~s integration. Visibility is normalized as in Figure \ref{fig:frrt_delay}.
The best estimate of instrumental delay has been removed for each curve.
}
\label{fig:delay_fine_structure}
\end{figure}

Along the delay axis, $|V(\tau, f)|$ takes the general form suggested by Figures\ \ref{fig:frrt_delay} and \ref{fig:delay_fine_structure}:
a narrow spike surrounded by a broad distribution.
We found that  the central spike takes the form of a sinc function in both delay and fringe rate coordinates,
as expected for uniform visibility across a square passband \citep{TMS2007}. 
The widths are somewhat larger than values expected from observing bandwidth of 16\ MHz and time span of 71.4\ s, of $w_{n\tau} =31.25$\ ns and $w_{nf}=14$\ mHz respectively,
probably because of the non-uniformity of receiver bandpasses and pulse-to-pulse intensity variations, respectively.
The broader part of the distribution takes an exponential form along the fringe-rate axis in this case;
more generally, the form can be complicated, particularly over times longer than 600\ s.
Traveling ionospheric disturbances may affect the time behavior of our 92-cm observations;
in particular, they may be responsible for the 20 to 25\ mHz width of the narrow component in fringe rate, as noted in Table\ \ref{tab:scatt_params}. 
We do not analyze the broader distribution in fringe rate further in this paper; we will discuss this distribution, and the influence of traveling ionospheric disturbances, in a separate publication (Paper III, Popov et al. in preparation). 
Because of the relatively small optical path length of the ionosphere, even at $\lambda=92$\ cm, they cannot affect the cross spectrum \citep{Hagfors1976}.

The distribution of the magnitude of the visibility in delay/fringe-rate domain changes with baseline length.
Figure~\ref{fig:delay_fine_structure} displays cross-sections through the maximum of the distribution of magnitude
for a range of baseline lengths, as a function of delay.
The maxima lie near zero fringe rate, as expected. Under the plausible and usual assumption that the correct fringe rate lies at the fringe rate, $f_{\rm max}$, where the distribution peaks,
the cross-section represents the visibility averaged over the time span of the sample:
\begin{equation}
V(\tau, f_{\rm max}) = \langle V(\tau, t) \rangle_{t}
\end{equation}
The top panel of Figure~\ref{fig:delay_fine_structure} shows this cross-section through Figure~\ref{fig:frrt_delay}.
The next lower panel shows the cross-section for the slightly shorter KL-RA baseline.
The three lower plots give the equivalent cross-sections for 10 times and 100 times shorter projected baselines. These three short-baseline 
cross-sections are qualitatively different from the long baseline cross sections: the visibility has a central spike
resulting from the component of the cross-spectrum that has a constant phase over frequency, as well as the 
broad distribution from the component that has a varying phase over frequency.
The central spike is strongest for the shortest baseline and weaker for the next longer baselines,
as expected based on the results of Sections\ \ref{sec:visibility} and\ \ref{sec:acf_visibility}. At very long
baselines the central spike is absent even after
averaging the visibility over the whole observing period, and only the broad component is present. 
As expected from Figure\ \ref{fig:frrt_delay},
in the delay/time domain the broad component appears as 
spikes distributed over a range of about $10~\mu$s in delay. These
spikes keep their position in delay for the scintillation time of
about 100 to 115~s, as listed in Table~\ref{tab:scatt_params}.

The character of the broad component changes with baseline length as well:
mean and mean square visibility are the same for short and long baselines; but excursions to small and large visibilities are more common for a long baseline 
\cite[][Eq. 12]{gwinn2001}.

\subsection{Averages and Correlation Functions}\label{sec:Obs_Corr_Func}

Averages of the visibility, and averages of the correlation function of visibility, extract the parameters of the broad and narrow components of visibility.
Such averages approximate the statistical averages discussed in Sections\ \ref{sec:Theory_Intro} and \ref{sec:impulse-response_visibility}.
They seek to reduce noise from the observing system and emission of the source,
as well as variations from the finite number of scintillations sampled, while preserving the statistics of scintillation.
The averages and correlation functions allow the inference of parameters of the impulse-response function of propagation
from the statistics of visibility.

\subsubsection{Square Modulus of Visibility $C$}\label{sec:corr_func_C}

The mean square modulus of visibility, 
$\langle C(\tau)\rangle_S = \langle |V(\tau)|^2 \rangle_S$,
provides useful and simple characterization of visibility.
To approximate the average over realizations of scattering $\langle...\rangle_S$, we average over many samples in time $t$
and over bins in delay $\tau$. 
We realize the average over time by evaluating $V(\tau, f)$ at the fringe rate of maximum amplitude $f_{max}$, as discussed in Section\ \ref{sec:Theory_Intro}.
We also average over 16 lags in delay $\tau$. 
The resulting average shows a broad component surrounding the origin; on shorter baselines, it shows a spike at the origin.
The broad distribution samples the properties of the fine structure seen in Figures\ \ref{fig:frrt_delay} and \ref{fig:delay_fine_structure},
and the spike to those seen on the shorter baselines in Figure \ref{fig:delay_fine_structure}.
We argue in Section\ \ref{sec:impulse-response_visibility} that the spike in $\langle C(\tau)\rangle_S$ is related to the average visibility,
and the broad component to the impulse-response function.

\begin{figure}
\centering
\includegraphics[angle=0,width=0.47\textwidth]{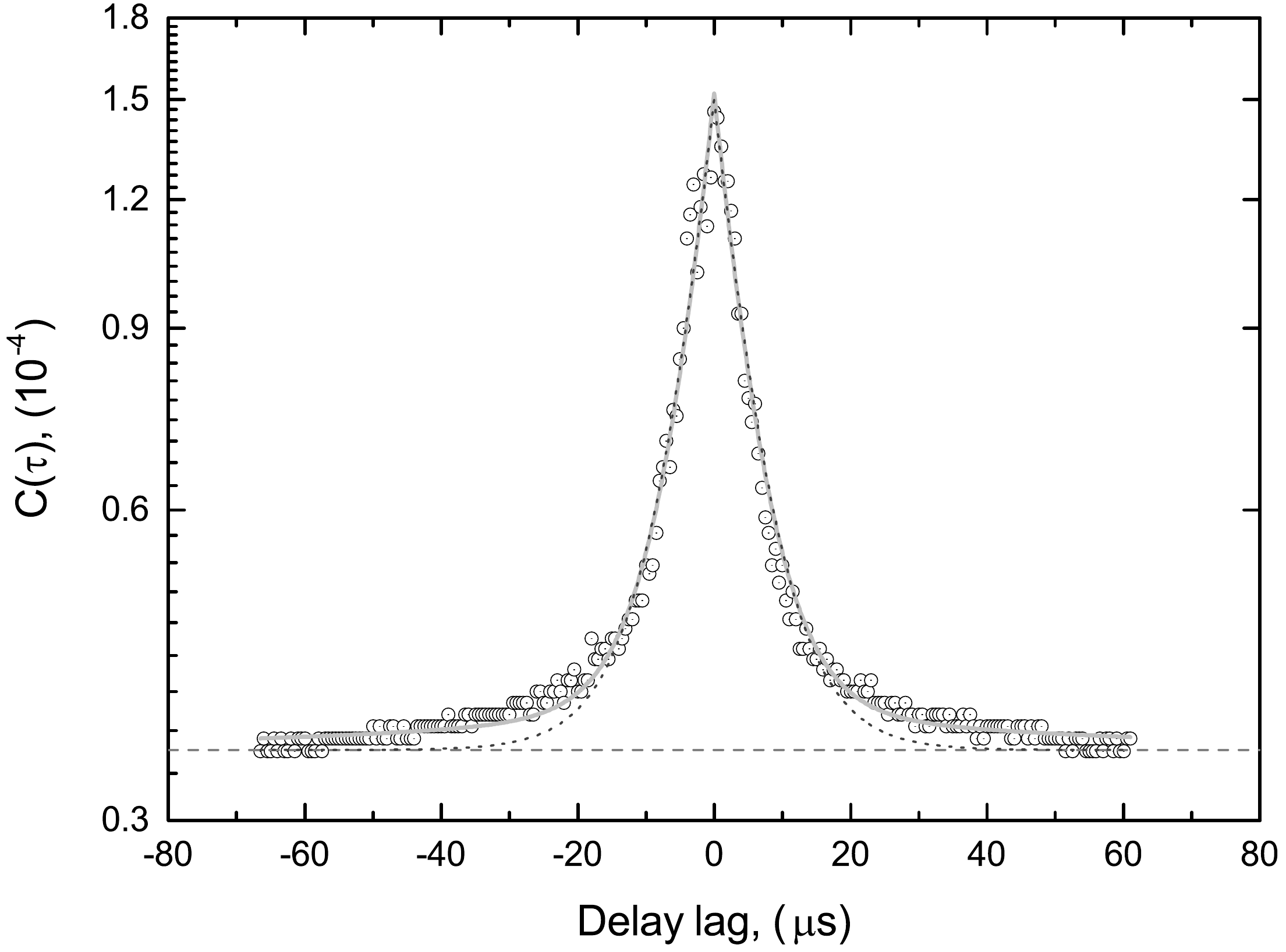}
\caption{Cross-section of the mean square visibility in the delay/fringe-rate domain $\langle C(\tau) \rangle_S =|V(\tau,f_{max})|^2$
along the delay axis, at the fringe rate $f_{max}$ where the magnitude of visibility peaks,  close to zero mHz. The visibilities for the GB-RA baseline on 2012 Nov 28 at 21:40 UT are shown as open circles.
The visibilities were computed by an inverse Fourier transform of the spectra, $\tilde V(\nu, t)$, over 71.4 s time spans, and then by averaging over 6 observing scans, each 570\ s long.
They were then further averaged in delay, over 16 points or $0.5$~$\mu$s, to smooth fluctuations. 
The dashed horizontal line shows the offset contributed by background noise. 
The solid gray line shows the reconstructed form given by Equation \ref{eq:Ctau_2exp_form}, offset by the noise level,
with parameters taken from the fit shown in Figure\ \ref{fig:KTwoScaleExamples}. The light dashed curve shows only the narrow component of the two-exponential model. Units of visibility are correlator units.}
\label{fig:Cross-section_FR-DL}
\end{figure}

Figure~\ref{fig:Cross-section_FR-DL} shows an example of the broad component of $\langle C(\tau)\rangle_S$.
This is estimated as $|V(\tau,f_{max})|^2$,
by selecting the peak fringe rate $f_{max}$ to average in time for each of 6 scans, averaging the results for the scans,
and averaging over 16 lags of delay to smooth the data.
These averaging procedures serve to approximate the average over an ensemble of realizations of scattering.
Background noise adds complex, zero-mean noise to $V(\tau,f)$, with uniform variance at all lags;
this adds a constant offset to the average $\langle C(\tau)\rangle_S = |V(\tau,f_{max})|^2$.

\subsubsection{Correlation Function $K$}

Using Equation\ \ref{eq:KRLdef}, we estimated $\langle K_{RL} (\Delta\tau )\rangle_S$,
the averaged cross-correlation function between the square modulus of right-circular polarized (RCP) and of left-circular polarized (LCP) of visibility in the
delay domain. 
(Note that $\langle K_{RL}(\Delta\tau)  \rangle_S$ is not the correlation function of the average $\langle C \rangle_S$, but rather the average of the correlation function
$\langle C_R \otimes C_L \rangle_S$.)
Because the background noise in the two circular polarizations is uncorrelated, they do not contribute an offset to $\langle K_{RL}(\Delta\tau)\rangle_S$.
This allows us to follow the effects of the impulse-response function to much lower levels than for $\langle C\rangle_S$.  
The correlation function $\langle K_{RL}(\Delta\tau)\rangle_S$ is thus less subject to effects of noise, and is more sensitive to the broad component of the distribution, than $\langle C\rangle_S$.

To compute an estimate of $\langle K_{RL} (\Delta\tau )\rangle_S$,
we calculated the squared sum of real and imaginary components of $V(\tau,t)$,
the inverse Fourier transform of the cross-power spectrum. We formed these for each
strong pulse, and normalized them by the autocorrelation functions at each antenna.
From these we formed the un-averaged correlation function $K_{RL}(\Delta\tau,t)$.
We then averaged $K_{RL}$ over 570-sec scans to form $\langle K_{RL} (\Delta\tau ) \rangle_{t}$.
Averaging in the time domain approximated an average over realizations of the impulse-response function for the scattering medium.
Each 570-sec scan included 100 to 250 strong pulses, yielding one averaged sample of $\langle K_{RL} (\Delta\tau,t ) \rangle_{t}$ for
each scan. We obtained 22 measurements in total, with 6
samples of $\langle K_{RL} (\Delta\tau,t)\rangle_{t}$  for November 26 , 28,  29
observing sessions. We obtained only 4 such samples for November 27
because of no significant detections of $V$ for two
scans on that date.

\begin{figure}
\centering
\includegraphics[angle=0,trim=1.8cm 1cm 2.1cm 1.5cm,width=0.5\textwidth]{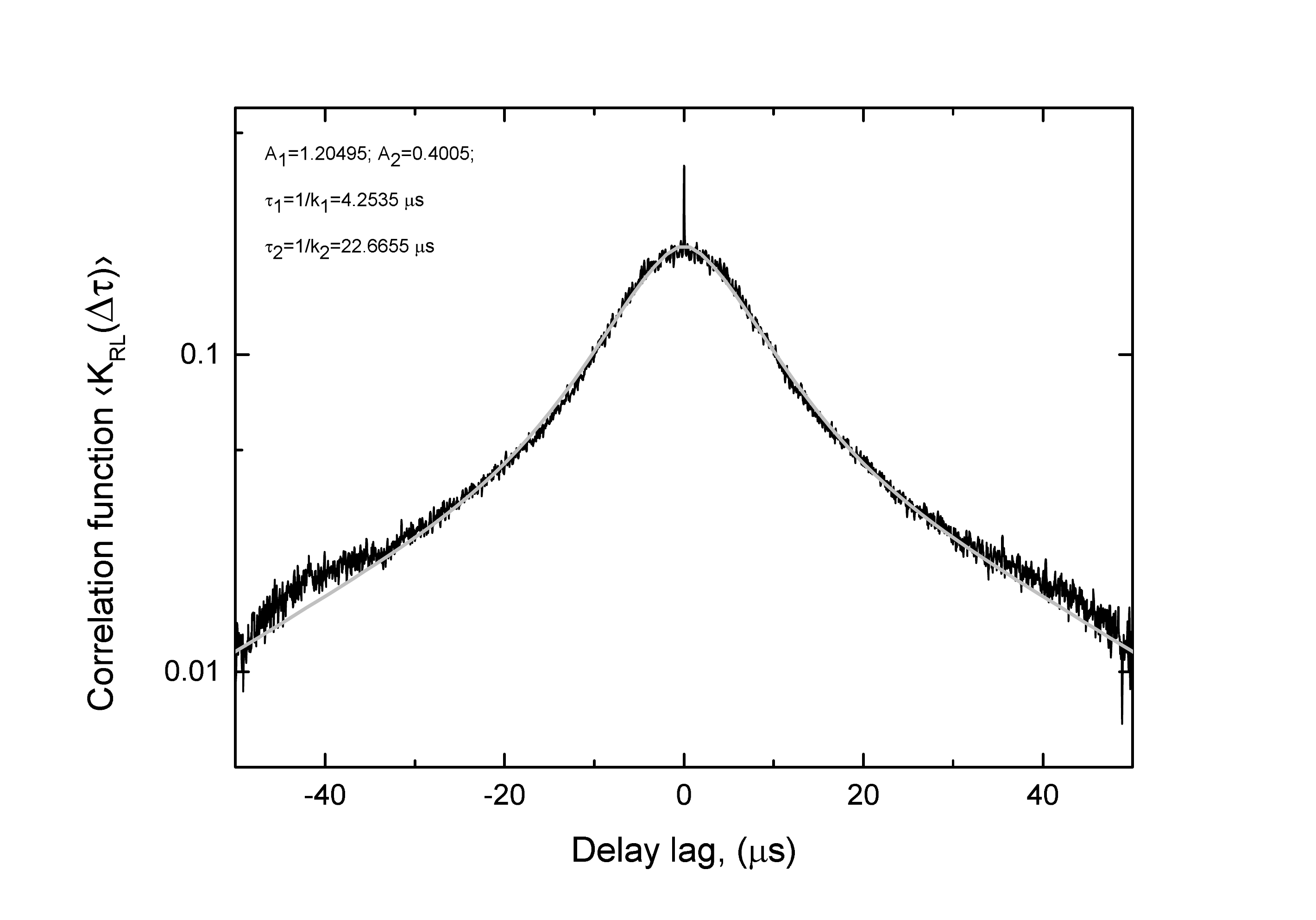}
\caption{An example of the correlation function $\langle K(\Delta\tau, f_{\rm max})\rangle_t$ on 2012 Nov 28,
averaged over 570\ s starting at 21:40:00 UT.
The data were normalized by the square root of $K_{RR}$ and $K_{RR}$ at $\Delta\tau=0$.
The best-fitting parameters for a 2-exponential fit of the form of Equation\ \ref{eq:K_2exponential_model} are as indicated.
 } \label{fig:KTwoScaleExamples}
\end{figure}

\subsubsection{Two Exponential Scales}\label{sec:two_scales}

Examination of the averaged cross correlation function, $\langle K_{RL}(\Delta\tau,t)\rangle_t$, revealed 
a spike at the origin and two exponential scales for the broad component,
a large one and a small one.
Figure\ \ref{fig:KTwoScaleExamples} shows an example.

The spike at the origin arises from the fine structure of scintillation in the broad component of visibility, 
as seen in Figures\ \ref{fig:frrt_delay} and\ \ref{fig:delay_fine_structure}.
This structure is identical in right- and left-circular polarizations,
so its correlation leads to the spike.

The two exponential scales are apparent as the slopes of the steeper and narrower parts of the distribution.
We see these two scales even for single pulses, which are strong enough to show the two-scale structure.
We did not observe these scales without doubt in spectra from single-dish observations, because the resulting correlation functions are more subject to noise, gain fluctuations, and interference.
The scales are both present for $\langle C\rangle_S$, but the longer scale is seen more clearly in $\langle K_{RL}(\Delta\tau)\rangle_S$ (as comparison of Equations\ \ref{eq:Ctau_2exp_form} through \ref{eq:K_2exponential_model_params} shows).

\subsubsection{Model Fit}

We formalized the two exponential scales seen for $\langle K(\Delta \tau)\rangle_S$ with a model fit.
The model assumed a pulse-broadening function with two exponential scales.
Under this assumption, a short pulse appears at the observer
with average shape:
\begin{equation}
G(\tau) = \begin{cases}
A_1 k_1 e^{-k_1 \tau} + A_2 k_2 e^{-k_2 \tau}, & \tau\ge 0 \\
0, & \tau<0
\end{cases}
\end{equation}
The pulse rises rapidly, and falls as the sum of the two exponentials.

The assumed form for $G$ leads to predictions for the forms of $\langle C\rangle$ and $\langle K\rangle$,
as discussed in Section\ \ref{sec:appendix2exponentials}.
For $\langle C\rangle$, we expect a cusp at the origin, and two exponentials with scales $k_1$ and $k_2$ and different weights
on either side.
For $\langle K\rangle$,
correlation smooths the cusp at the origin, producing a smooth peak, with the same exponential scales appearing to either side.

Figure\ \ref{fig:KTwoScaleExamples} shows the best-fitting model of this form for the data shown there.
This model has parameters:
\begin{align}
A_1/A_2 &= 0.33 \label{eq:2expKfitparams}\\
k_1 &=1/4.3\ \mu{\rm s}\\
k_2 &= 1/23\ \mu{\rm s}.
\end{align}
The model reproduces the two scales, and the smooth peak, well.  The model also predicts the magnitude of the spike accurately,
with zero average visibility $\rho_{AB}=0$.

The model shown in Figure\ \ref{fig:Cross-section_FR-DL} shows the model for $C$, reconstructed using Equation\ \ref{eq:Ctau_2exp_form} with parameters from the fit to Figure\ \ref{fig:KTwoScaleExamples}.
The two scales appear in the model, although the offset from noise contributes at large delay $\tau$.
As the figure shows, a single exponential does not fit the model well: the narrow component is satisfactory at small $\tau$, but falls well under the data at larger $\tau$.
A high-winged function such as a Lorentzian can fit $C$ well, but the rounded peak leads to a very wide peak for $K$ that cannot match the data, and the inversion to a $G(t)$ that remains finite, and is zero for $t<0$
as causality demands, is problematic.

\begin{figure}
\centering
\includegraphics[angle=0,trim=2cm 0.5cm 2.5cm 0.5cm,width=0.5\textwidth]{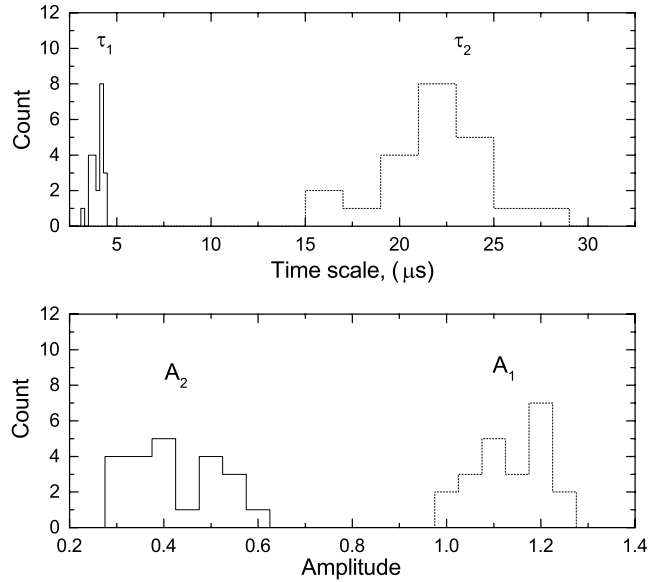}
\caption{Upper panel: The distribution of long and short time scales for exponential scales of $\langle K_{RL}(\Delta\tau, f_{\rm max})\rangle_t$. 
Each pair of scales was measured for a 570-s interval on one of the 4 consecutive observing days in 2012.
Lower panel: The distribution of magnitudes of long and short time scales.} \label{fig:ScaleDistributions}
\end{figure}

The best-fitting scales and the magnitudes of the two contributions varied from scan to scan, but in a manner that was consistent with our finite 
sample of the scintillation pattern, and the inhomogeneous averaging of pulses with different intensities.
We show a histogram of the results of our fits to 570-sec intervals in Figure\ \ref{fig:ScaleDistributions}.
On 26 to 29 November 2012, the shorter scale averaged to $\tau_1 =4.1\pm0.3\ \mu{\rm s}$, and the longer scale to $\tau_2 = 22.5\pm 2.9\ \mu{\rm s}$. 
The scales had a relative power of $A_2/A_1 = 0.38$.

\section{Discussion}\label{sec:discussion}

On a long baseline that fully resolves the scattering disk,
as Figure\ \ref{fig:delay_fine_structure} shows,
we observe multiple sharp spikes in the visibility $V(\tau, f)$ is a consequence of the variation of 
the amplitude and phase of visibility.
(See also Paper II, Popov et al. in preparation).
The characteristic region of that variation,
$\Delta\tau \cdot \Delta f$, reflects the product of the 
inverses of the scintillation bandwidth $\Delta \tau \approx 1/2\pi \Delta\nu_{\rm sc}$ and the scintillation timescale $\Delta f \approx 1/2\pi t_{\rm sc}$.
These quantities are the width in time of the impulse-response function,
and the time for the impulse-response function to change as the line of sight to the observer moves through the scattering material.

Detailed examination of the correlation function of visibility $K_{RL}(\Delta \tau,t)$ reveals the presence of two characteristic, exponential scales. 
Both scales are visible in the single-pulse correlation functions of right and left circular polarization,
as well as in the correlation function $\langle K_{RL} (\Delta\tau)\rangle_t$ averaged over 570\ s shown in Figure\ \ref{fig:KTwoScaleExamples}.
For an assumed screen distance of half the pulsar distance of $D=1.03^{+0.13}_{-0.12}$~kpc \citep{brisken2002},
the two scales correspond to diffractive scales of: 
\begin{align}  
\ell_{d1} &= \frac{\lambda}{2\pi} \sqrt{ \frac{ D} {c\tau_1}}=2.3 \times 10^{9}\ {\rm cm} \\
\ell_{d2} &=\frac{\lambda}{2\pi} \sqrt{ \frac{ D} {c\tau_2}} =1.0 \times 10^{9}\ {\rm cm} \nonumber
\end{align}
The diffractive scale is the lateral distance at the screen where phases decorrelate by a radian \citep{narayan1992}.
The refractive scale gives the scale of the scattering disk:
\begin{align}  
\ell_{r1} &= \sqrt{ c\tau_1 D} = 1.9\times 10^{13}\ {\rm cm} \\
\ell_{r2} &= \sqrt{ c\tau_2 D} = 4.6\times 10^{13}\ {\rm cm} \nonumber
\end{align}
In contrast, \citet{britton1998} measured angular broadening for \psr\ of $\theta_H < 3.4$\ mas at $\nu=325$\ MHz,
where $\theta_H$ is the full width of the scattered image at half the maxium intensity.
This corresponds to a refractive scale of $\ell_r = (\theta_H/\sqrt{8 \ln 2}) D/2 < 1.1\times 10^{13}\ {\rm cm}$.  
This upper limit is somewhat smaller than the values obtained from our observations, even if one takes into account the facts that the larger scale contains only 0.38 of the power of the shorter one,
and that the scattering material may be somewhat closer to the pulsar than to the observer.

\subsection{Previous Observations}

\citet{shishov2003} studied the scattering properties of \psr in detail,
using single-antenna observations at 102~MHz, 610~MHz, 5~GHz, and 10.6~GHz to form structure functions of the scintillation in time and frequency on a wide range of scales.
They concluded that the scattering material has a power-law spatial spectrum with index $\alpha+2=3.50\pm 0.05$, marginally consistent with the value of $11/3$ expected for a Kolmogorov spectrum,
with an outer scale of $2\times 10^{11}\ {\rm m} < L_0 < 10^{17}\ {\rm m}$.
Using VLBI,
\citet{bartel1985} observed \psr at 2.3\ GHz and set limits on the separation on the emission regions corresponding to different components of the pulse profile.
\citet{yangalov2001} observed \psr at $1.650$\ GHz with ground-space baselines to HALCA,
and found that the source varied strongly with time.
They ascribed this variation to scintillation, with the scintillation bandwidth comparable to the observing bandwidth at their observing frequency.
Self-calibration with timespans less than the scintillation time returned a pointlike image, as expected.
\citet{semenkov2004} analyzed these data, including ground-ground baselines. They studied both single-antenna autocorrelation functions $V_{AA}(\Delta\tau)$ and cross-correlation functions $V_{AB}(\Delta \tau)$.
They detected two timescales for the scintillation pattern,
of 20\ min and 1\ min.
They found that the properties of scattering could not be explained by a single, thin screen, and further
that velocities indicated relative motions within the scattering medium.

\citet{popov1984} had previously observed two coexisting scales of scattering for 
\psr. 
They found scintillation bandwidths of $\Delta \nu_1 = 115$\ Hz and $\Delta\nu_2 = 750$\ Hz,
measured as the $1/e$ point of the correlation function of intensity at an observing frequency of 102\ MHz,
using the Large Cophase Array of Puschino Observatory.
The ratio of these scales, $\Delta\nu_2/\Delta\nu_1=6.5$ is larger than the ratio of $k_1/k_2=5.5$ that we observe.  
Scaled to our observing frequency of 324\ MHz, using the $\Delta\nu \propto \nu^{22/5}$ scaling appropriate for a Kolmogorov spectrum,
and converting from $\Delta\nu$ to $\tau$ using the uncertainty relation $\tau = 1/2\pi \Delta\nu$,
we find that these values
correspond to 1.3 and 8\ $\mu$s, respectively, about a factor of 3 smaller that the scales we observe.  
Of course, interpolation over a factor of 3 in observing frequency and the different observing techniques may introduce biases,
and scattering parameters likely vary over the years between the two measurements.
Two scales of scattering have also been observed for other pulsars \citep{gwinn2006,smirnova2014}.

\subsection{Origin of Two Scales}

Two scales of scattering may be a consequence of a variety of factors.
Non-Gaussian statistics of scattering can produce multiple scales,
although this usually appears as a continuum of scales rather than two different individual scales,
as in a power-law distribution or a Levy flight \citep{boldyrev2003}.
A Kolmogorov model for scattering in a thin screen does not fit as well as our model based upon a two-exponential impulse-response function, or even as one based upon one exponential.
A model with two discrete scales appears to fit our data better.

One explanation is anisotropic scattering.  This can produce two scales, corresponding to the major and minor axes of the scattering disk,
as discussed in Section\ \ref{sec:anisotropic} of the Appendix.
The ratio of the scales of $k_2/k_1 = 5.5$ corresponds to the parameter $\alpha^2=57$, 
and an axial ratio of $\theta_2/\theta_1 =\sqrt{ 1+ \alpha^2} =\sqrt{2(k_2/k_1)^2 -3}= 7.4$.  
In a simple model for anisotropic scattering in a thin screen, we expect the ratio of power in the scales to be approximately 
$\sqrt{ {1+\alpha^2}/{2\pi (2+\alpha^2)}  }\approx 0.40$, as shown in Section\ \ref{sec:anisotropic} of the Appendix.
This compares well with our observed ratio of $A_1/A_2=0.38$.
However, our observations for ground-space baselines at a variety of orientations do not show anisotropy.
A variety of models, involving material with varying anisotropy distributed along the line of sight, and strong anisotropy that slips between our long baselines, might match our data.

A second explanation is the complicated structure observed within dynamic spectra: most commonly observed as ``scintillation arcs'' \citep{stinebring2001}.
Recently, it has been suggested that this structure arises from
interference among subimages, resulting from refraction by interstellar reconnection sheets \citep{PenLevin2014}.
This complicated structure produces time and frequency variations on a wide range of scales.
Of course, we are considering very long baselines, where the scintillation-arc patterns should be completely uncorrelated between antennas.  This may lead to 
blurring, resulting in a 2-scale correlation function without particularly strong structure corresponding to the discrete arcs seen on shorter baselines \citep{brisken2010}.
We do not see any direct evidence of scintillation arcs, as such.  The magnitude of the visibility shows a featureless decline with increase of either of the 2 dimensions $|\tau|$ and $|f|$. 
The GB autocorrelation functions do not show scintillation arcs either, for our observations. 

\section{Summary}\label{sec:summary}

We made VLBI observations of \psr with \RA at 324 MHz on projected baselines of up to 235,000~km. Our goal was to investigate scattering by the interstellar medium. These properties affect radio observations of all celestial sources. While the results of such observations are in general influenced by the convolution of source structure with the scattering processes,  pulsars are virtually point-like sources and signatures in the observational results can be directly related to the scattering properties of the interstellar medium.

On long baselines, in the domain of delay $\tau$ and fringe-rate $f$, 
the correlation function of visibility $V(\tau,f)$ is a collection of narrow spikes, 
located within a region defined by the inverses of the scintillation bandwidth $\Delta \tau \approx 1/2\pi \Delta\nu_{\rm sc}$ and the scintillation timescale $\Delta f \approx 1/2\pi t_{\rm sc}$.
For shorter baselines, a sharp spike at the center of this region represents the average visibility;
on long baselines where the average visibility drops to near zero, this spike is absent.

The mean square visibility, $\langle C(\tau)\rangle_S = \langle |V(\tau)|^2\rangle_S$, is well fit with a smooth model, indicating that the visibility spikes are the result of random interference of many scattered rays. 
To form a quantity less subject to effects of noise, 
we convolve the mean-square left- and right-circular polarized visibility to form
$\langle K_{RL}(\Delta\tau)\rangle_S = \langle |V_L(\tau)|^2   \convolve_{\tau\rightarrow\Delta\tau} |V_R(\tau)|^2 \rangle_S$.
The average correlation function $\langle K_{RL}(\Delta\tau)\rangle_S$
shows two exponentials with different characteristic timescales.  
The forms of $\langle C(\tau)\rangle$ and $\langle K_{RL}(\tau)\rangle$ are well fit with a simple model, that assumes that the average pulse-broadening function $G$ is the sum of two exponentials with different timescales.

On 2012 Nov 26 to 29, the shorter timescale was $4.1\pm 0.3\ \mu{\rm s}$, and the longer timescale was $23\pm 3 \mu{\rm s}$, with the longer-scale exponential containing approximately 0.38 times the power of the shorter-scale exponential. 
This double exponential may arise from anisotropic scattering; or from 
scattered radiation at large angle, perhaps corresponding to the subimages seen in single-dish and shorter-baseline observations.
Further investigation of the properties of the image of the scattered pulsar on long and short baselines, using these data, will help to clarify the origin of the two scales.

\acknowledgments
The RadioAstron project is led by the Astro Space Center of the Lebedev Physical Institute of the Russian Academy of Sciences and the Lavochkin Scientific and Production Association under a contract with the Russian Federal Space Agency, in collaboration with partner organizations in Russia and other countries. The National Radio Astronomy Observatory is a facility of the National Science Foundation operated under cooperative agreement by Associated Universities, Inc. This study was supported by Russian Foundation for Basic Research grant 13-02-00460 and Basic Research Program P-7 of the Presidium of the Russian Academy of Sciences.  C.R.G.\ acknowledges support of the US National Science Foundation (AST-1008865). N.B.\ was supported by NSERC.
We thank the referee for constructive comments which helped to improve the manuscript.

{\it Facilities:}
\facility{\textit{RadioAstron} Space Radio Telescope (Spektr-R), GBT, WSRT, Kalyazin radio telescope}.

\appendix

\section{Impulse-Response Function and Visibility}\label{sec:impulse-response_visibility}

\subsection{Introduction}

Under general assumptions, refraction and scattering convolve the electric field of a source with an impulse-response function $g$.
This function varies with position in the observer plane, decorrelating over some lateral scale; and with time, as the line of sight to the source moves with respect to the scattering material,
and as the scattering material evolves.
The task of this section is to relate the impulse-response function to the statistics of visibility, as given by the functions $C$ and $K$ introduced in Section\ \ref{sec:Theory_Intro} above.

\subsubsection{Notation}

The visibility $V_{AB}$ is the conjugated product of electric fields at two antennas (Equation\ \ref{eq:Vdef}). We usually omit the subscripts indicating baseline $AB$ on $V$, unless they are important for the immediate argument.
We denote the Fourier transform from the time or delay domain ($t$ or $\tau$) to the frequency or fringe rate domain ($\nu$ or $f$) by ${\mathfrak F}$, and its inverse by ${\mathfrak F}^{-1}$.
We accent symbols with tilde ``$\tilde\ $'' to denote quantities that depend on observing frequency $\nu$,
and the same symbols without accent for the Fourier-conjugate domain of delay $\tau$ or time $t_e$.
We assume that the variables 
describing 
time and frequency $t_e, \nu, \tau, t, f$, are discrete.
They range from $-N/2$ to $N/2-1$, where $N$ is the number of samples in the time or frequency span.
For $\tau$ and $t_e$ one sample is the inverse of the Nyquist rate, and they can span the time to accumulate a single realization of the spectrum;
for $t$ and its Fourier conjugate $f$ one sample is the averaging time for one spectrum, and they can span one observation.

Our convention for normalization of the Fourier transform is that a function $h(\tau)$ 
normalized to unit area in the delay domain has value unity at zero frequency: $\tilde h(\nu=0) =1$.
Conversely, if $\tilde h(\nu)$ is normalized to unit area in the frequency domain, $h(\tau=0)=1/N$.
This is the ``$\{ 1, -1 \}$'' convention of Wolfram Mathematica \citep{FTConvention}.
With this convention, Parseval's Theorem takes the form:
\begin{align}
\sum_{\tau=-N/2}^{N/2-1} h(\tau) h^*(\tau) &=\frac{1}{N}\sum_{\nu=-N/2}^{N/2-1} \tilde h(\nu) \tilde h^*(\nu) 
\end{align}

\subsection{Impulse-Response Function}\label{sec:AppImpluseResponseFunction}

As noted above, the observed electric field
of a pulsar $E_{obs}(t_o)$ is 
the convolution of the electric field emitted at the source with a kernel $g$ that depends on scattering:
\begin{align}
E_{obs} (t_o) &= \sum_{t_e=0}^{{\rm few}\times \tau_{\rm sc}} g(t_e) \,E_{src}(t_o-t_e) = g\otimes E_{src}
\label{eq:g_conv}
\end{align}
where we introduce the symbol $\otimes$ for convolution.
The kernel $g$ is the impulse-response function;
in other words, if the pulsar emits a sharp spike, then the observed electric field of the pulse is simply a copy of $g$.
Because of this convolution, $g$ is also known as the propagation kernel; it is also known as the Green's function, and the S matrix
\citep[][and references therein]{GwinnJohnson2011}.
Both $E_{src}$ and $g$ vary at the Nyquist rate: the inverse of the total observed bandwidth.
Usually, we assume that the intrinsic electric field of the source is white noise at the Nyquist rate: it is drawn from a Gaussian distribution in the complex plane at each instant \citep{rickett1975}. 
The impulse-response function extends over a time span of a few times $\tau_{\rm sc}$, representing the time over which a sharp pulse at the source would be received.
It is zero outside this relatively narrow time window.

If the statistics of the scattering material are stationary,
the characteristic shape and scales of $g$ will remain fixed, while details of amplitude and phase vary
on the timescale $t_{\rm sc}$ (Equation \ref{eq:t_sc}).
An average of the squared electric field over many impulses emitted by the source over times longer than $t_{\rm sc}$ will reveal the characteristic form.
One simple model form for $g$ that includes deterministic and random parts is the product of a non-varying envelope $g_D(t_e)$, 
and a random function $g_R$, that varies rapidly with $t_e$ during the course of each pulse:
\begin{align}
g (t_e) &= g_D (t_e)\cdot g_R (t_e, t) 
\label{eq:gR_gD_conv}
\end{align}
Both $g_R$ and $g_D$ span a few times $\tau_{\rm sc}$. 
Over that time span, $g_R$ varies wildly, and randomly; however, it exhibits nearly the same form for the next pulse.
Over the longer timescale $t_{\rm sc} \gg \tau_{\rm sc}$, 
the form of the random function $g_R$ changes slowly.
For typical observations of a pulsar, such as those described in this paper, $\tau_{\rm sc}$ is some fraction of the width of one pulse, or a few microseconds; whereas $t_{\rm sc}$
is many pulsar periods, or many seconds.
Such situations, where a convolution may have a slowly-varying kernel, are commonly treated as ``dynamic spectra'' \citep[see][Ch. 19]{bracewell2000}.

The intensity received by an observer
for an electric-field impulse at the source, 
averaged over many such impulses with different realizations of the scattering material,
is the square modulus of the deterministic part of $g$, which we call $G$:
\begin{align}
\Big\langle I_{obs}(t_e) \Big\rangle_S\equiv G(t_e) &= g_D(t_e)\cdot g_D^*(t_e)
\label{eq:Gdef}
\end{align}
Here, the subscripted angular brackets $\langle ... \rangle_{S}$ indicate a statistical average over realizations of the scattering medium, 
for example as approximated by an average over pulses spanning a time greater than $t_{\rm sc}$.
Often, $G$ is called the pulse-broadening function.
So that propagation kernel leaves the intensity of the source unchanged, when averaged over time, we set:
\begin{align}
\sum_{t_e} G(t_e) &\equiv 1
\label{eq:normalize_gD}
\end{align}

For strong scattering, as is observed for most pulsars at most wavelengths,
we expect that many different paths,
with random amplitude and phase, will contribute to the received pulse at each instant $t_e$.
Therefore, we expect that the random part $g_R$ will have the statistics of a random walk at each instant:
the observed electric field will be drawn from a circularly Gaussian distribution in the complex plane,
with zero mean.
On the other hand, the deterministic part $g_D$ sets the standard deviation of $g$;
it reflects how many paths, and with what strength, contribute at each delay.
This model for scintillation is closely related to the amplitude-modulated-noise (AMN) model for pulsar emission \citep{rickett1975}.
In this model, the electric field emitted by the pulsar is the product of noise, drawn from a zero-mean Gaussian distribution in the complex plane,
with a more-slowly varying envelope that determines the standard deviation of the noise at each instant.

We suppose that the random part of the propagation kernel is completely uncorrelated in time, at the Nyquist rate, within its span of a few $\tau_{\rm sc}$.
Then, at a location ``$A$,''
\begin{align}
\big\langle g_{RA}(t_e ,t ) g_{RA}^*( t_{e}+\tau , t ) \big\rangle_{S} &=
\begin{cases}
 1 & {\rm if\ } \tau=0 \\
 0 & {\rm if\ } \tau\neq 0
\end{cases}
\label{eq:normalize_gRA}
\end{align}
On the other hand, $g_{RA}$ is nearly the same for each emitted pulse; it changes only over the longer timescale $t_{\rm sc}$.
The question of how this slower variation of $g_R$ with time depends upon baseline length is much more complicated,
and we discuss it briefly below.
However, if the lateral separation of the two stations $A$ and $B$ is much greater than the scale of the scattering pattern,
then the random parts of $g$ for the two stations, $g_{RA}$ and $g_{RB}$, are completely uncorrelated.

\subsection{Visibility: Dynamic Cross-Power Spectrum}\label{sec:def_dynamic_cross_spectrum}

As the previous discussion shows, the impulse-response function involves three timescales:
$g_R$ changes at the Nyquist rate;
$g_D$ varies over the typical span of the impulse-response function $\tau_{\rm sc}$;
and the time for the random variations of $g_R$ to change is $t_{\rm sc}$.
The dynamic cross-power spectrum provides a useful description for these different variations \citep{bracewell2000}.
A single sample of the cross-power spectrum, when averaged over time less than $t_{\rm sc}$, has the characteristic scale $\Delta \nu_{\rm sc}\approx 1/2\pi \tau_{\rm sc}$,
resulting from the finite span of the impulse-response function and the uncertainty principle.
The time variation of the cross-power spectrum over times $t\geq t_{\rm sc}$ captures the changes of $g_R$.

Visibility $\tilde V$ in the domain of frequency $\nu$ and time $t$
is the product of the Fourier transforms of electric fields at stations A and B (see Equation\ \ref{eq:Vdef}):
\begin{align}
\tilde V_{AB} (\nu, t) &= \tilde E_{A}(\nu,t) \tilde E_{B}^* (\nu,t) 
\label{eq:tildeV_def}
\end{align}
We suppose that each sample of the cross-spectrum is averaged over many realizations of the source electric field $E_{src}$, 
over a time short compared with $t_{\rm sc}$.
This reduces noise from the source and backgrounds.

One may represent the visibility in four domains, linked by Fourier transforms of frequency $\nu$ to delay $\tau$, and time $t$ to fringe rate $f$, as Figure\ \ref{fig:FTCircle} illustrates.
In this paper we are particularly concerned with $V_{AB}(\tau,t)$, visibility in the domain of delay $\tau$ and time $t$.
This is the Fourier transform of $\tilde V_{AB} (\nu, t)$. The convolution theorem for Fourier transforms shows that $V(\tau,t)$ is the cross-correlation function of electric fields
in the time domain:
\begin{align}
V_{AB} (\tau ,t) &=\sum_{t_e=-N/2}^{N/2-1} E_{A}(t+t_e) E_{B}^* (t+t_e+\tau) 
\label{eq:Vtau_def}
\end{align}
Here, $t_e$ indexes individual samples of electric field, over a short interval near the index time of the measurement of the cross-power spectrum, $t$.

Visibility in the delay-rate domain, conjugate to the frequency-time domain, takes the form:
\begin{align}
V (\tau, f) &= {\mathfrak F}^{-1}_{\nu\rightarrow \tau}\left[  {\mathfrak F}_{t\rightarrow f}\left[ \tilde V (\nu, t) \right] \right]
\label{eq:VtauFTVnu}
\end{align}
Searches for interference fringes are often conducted in this domain:
because absolute calibration of delay and fringe rate are usually impossible for very-long baseline interferometry,
the peak of $|V(\tau,f)|$ can be used to determine them \citep{TMS2007}.

\subsubsection{Visibility and Impulse-Response Function}\label{sec:visibility}

Visibility depends on the separation $\vec b$ of stations $A$ and $B$, as well as on delay and rate, or time and frequency.
From Equations\ \ref{eq:g_conv} and\ \ref{eq:Vtau_def}, and the assumption that the electric field of the source $E_{src}$ is a stationary random variable without correlation in time, we
find that
visibility in the delay domain is the cross-correlation function of $g$ at the two stations:
\begin{align}
V_{AB} (\tau) & = g_A(\kappa) \convolve_{\kappa\rightarrow \tau} g_B(-\kappa)
\label{eq:VABtau_as_g}
\end{align}
This leads to the expected form of $V(\tau)$: a spike at $\tau=0$, with average magnitude equal to the average correlation $\rho_{AB}$;
and a broad component of width $\tau_{\rm sc}$, with random amplitude, and phase variations that increase with baseline length, corresponding to the random character of $g_R$ and its decorrelation with increasing baseline.
Equations\ \ref{eq:normalize_gRA} and\ \ref{eq:normalize_gD} show that
$\rho_{AB}=1$ for $g_{RA}=g_{RB}$,
and 
$\rho_{AB} \rightarrow 0$ for uncorrelated $g_{RA}$ and $g_{RB}$.
Figures\ \ref{fig:frrt_delay} and \ref{fig:delay_fine_structure} show examples.

On intermediate baselines,  the time structure of the correlation of $g_R$ is more complicated,
in a way that depends on the geometrical distribution of the paths that contribute to $g_R$.
For scattering material concentrated in a thin screen, for example,
the shortest-length paths result in small $t_e$ in the impulse-response function, and also tend to appear at small angles at the observer.
Thus, at small delays correlation is high even for rather long baselines; whereas at long delays correlation is poor even for shorter baselines.
Thus, correlation between antennas should decrease at later times $t_e$ within $g_R$.
This correlation is imprecise for scattering material distributed along the line of sight, where many deflections along the line of sight lead to a large time lag $t_e$, but little or no angular deflection at the observer's interferometer.
Moreover, in the frequency domain, dynamic single-dish spectra can show slants and complicated patterns \citep{hewish1980,stinebring2001},
suggesting complicated correlations of time and delay in the observer plane.
Analysis of the visibility is thus easiest on very short baselines and very long ones.

Equations\ \ref{eq:VtauFTVnu} and\ \ref{eq:VABtau_as_g} provide the relation of the visibility in the frequency-time domain 
to the impulse-response function:
\begin{align}
\tilde V_{AB} &  = 
\left( \tilde g_D \otimes \tilde g_{RA} \right) \left( \tilde g^*_D \otimes \tilde g^*_{RB} \right) 
\label{eq:VAB_as_g}
\end{align}
Averaging the visibility over many scintillations yields the average visibility, $\rho_{AB}$:
\begin{align}
\left\langle \tilde V_{AB} \right\rangle_S &  \equiv  \rho_{AB} 
\end{align}

\subsubsection{Square Modulus of Visibility $C$ and Correlation Function $K$}\label{sec:acf_visibility}

The average of $V(\tau)$ over many realizations of scintillations leaves the delta-function at the origin that corresponds to the average visibility:
\begin{align}
\Big\langle V_{AB} (\tau) \Big\rangle_S & =0 + 
 \rho_{AB} \, \delta^0_\tau
\end{align}
Here, $\delta^\tau_0$ is the Kronecker delta-function, with value 1 if $\tau=0$ and 0 otherwise.

The secondary spectrum $C(\tau,f)$ may be defined as the square modulus of $V(\tau,f)$\footnote{\citet{brisken2010} 
define the secondary spectrum as $V (\tau,f) V(-\tau,-f)$. For zero baselines $\tilde V(\nu,t)$ is real, 
so that $V (\tau,f)=V^*(-\tau,-f)$, and our and their expressions are identical. 
Their expression includes phase information in an elegant way for their short baseline, where departures from zero phase are small. 
For observations on long baselines, their expression is impractical because identification of the origin of $(\tau,f)$ is not possible, as Figure\ \ref{fig:frrt_delay} shows. 
Consequently, the pair $(\tau,f)$ and $(-\tau,-f)$ cannot be combined reliably.
They also use the accent $\tilde\ $ for visibility in the delay domain, although not for $C$; whereas we use the accent for quantities in the frequency domain of $\nu$.
}:
\begin{align}
C (\tau,f) &= V (\tau,f) V^* (\tau,f) = |V(\tau,f)|^2
\end{align}
This function provides information similar to $|V(\tau,f)|$, shown in Figure\ \ref{fig:frrt_delay}, but is easier to deal with statistically.
Because $C (\tau,f)$ is the square modulus of the complex visibility $V(\tau, f)$, background noise adds noise to $C$, except for the central lag, where it may contribute a constant offset.  
Similarly, self-noise will add noise with an envelope that follows the average form of $C$. This behavior is in contrast to that of correlation functions of single-dish quantities such as the intensity, where noise can contribute to the mean correlation function.

Our long baselines fully resolve the scattered image. The phases of scintillation elements in $\tilde V(\nu,t)$ appear to be random, and the phases of $C(\tau,f)$ show no discernible patterns.
An inverse Fourier transform from fringe rate $f$ to time $t$ leads to $C(\tau,t)$:
\begin{align}
C (\tau,t) &= {\mathfrak F}^{-1}_{f\rightarrow t} \left[ C(\tau,f) \right] .
\end{align}
Evaluated at the fringe rate $f_{max}$ of its peak magnitude, the secondary spectrum $C(\tau,f_{max})$ is a time average, that
approximates an average over realizations of the scintillation pattern of $C(\tau,t)$:
\begin{align}
\Big\langle C (\tau) \Big\rangle_S &  \approx C(\tau,f_{max})
\end{align}
The autocorrelation function of $C$ is $K$:
\begin{align}
  K (\Delta\tau ) &= C(\tau,t) C(\tau+\Delta \tau,t)   \label{eq:Ktau_def} 
\end{align}
Note that $C(\tau,t)$ appears without averaging in this expression.
Conveniently, the correlation $K_{RL}(\Delta\tau, t)$ between the secondary spectra $C$ in the right and left circular polarizations eliminates some effects of noise
and interference,
as noted in Section~\ref{sec:two_scales}.

The behaviors of $C$ and $K$ are simplest to describe on very short baselines, where $g_A(\kappa) = g_B(\kappa)$ for all $\kappa$, and $\rho_{AB}=1$;
and for very long baselines, where the random parts of $g_A(\kappa)$ and $g_B(\kappa)$ are completely uncorrelated, so that $\rho_{AB}=0$.
If the correlation of the random parts of the propagation kernels $\rho_{AB}$ is constant, then the ensemble-average values of these correlation functions are:
\begin{align}
\Big\langle C (\tau) \Big\rangle_S  &=   \frac{ C_2 }{N} \big( G \otimes G_- \big) + \left(\rho_{AB}^2\right) \delta^\tau_0 
\label{eq:C_result}\\
\Big\langle K (\Delta\tau ) \Big\rangle_S &=
\frac{ C_2 }{N}  \big(G \otimes G_-  \otimes G \otimes G_-  \big) ( 1 + 4 \rho_{AB}^2 ) + 2 \left( \frac{C_2 }{N} +  \rho_{AB}^4 \right)\delta^{\Delta\tau}_0
\label{eq:K_result}
\end{align}
Again, $\delta^\tau_0$ is the Kronecker delta-function, with value 1 if $\tau=0$ and 0 otherwise; and similarly for $\Delta \tau$.
The time-reversed pulse-broadening function is $G_-$, given by $G_-(\kappa) = G(-\kappa)$.
The constant $C_2$ is the mean square of $G$: $C_2 = \sum_{t_e} G^2$.

\section{Two Exponentials}\label{sec:appendix2exponentials}

We consider a situation where the impulse-response function is the sum of two exponentials with different time constants $k_1$ and $k_2$,
with a rapid rise from $G=0$ at $t=0$.
As we discuss below, this may result in a variety of circumstances.
We parametrize the impulse-response function:
\begin{equation}
G(\tau) = 
\begin{cases} 
A_1 k_1 e^{-k_1 \tau} + A_2 k_2 e^{-k_2 \tau} & \tau\ge 0 \\
0 & \tau<0
\end{cases}
\end{equation}
In the text, we also make use of the inverse scales $\tau_1=1/k_1$ and $\tau_2=1/k2$; these can provide better physical insight.
The autocorrelation of $G$ provides the form for $\langle C(\tau) \rangle_S $, the mean square visibility in the delay-time domain, as given by Equation\ \ref{eq:C_result}.
For this impulse-response function, under the assumption that the baseline is so long that $\rho_{AB}=0$, this takes the form:
\begin{equation}
\frac{N}{C_2 } \Big\langle C(\Delta \tau)\Big\rangle_S = A_1 k_1 \left( \frac{A_2 k_2}{k_1+k_2}+\frac{A_1}{2}\right) e^{-k_1 \left| \Delta \tau \right|} 
+  A_2 k_2 \left( \frac{A_1 k_1}{k_1+k_2}+\frac{A_2}{2}\right) e^{-k_2 \left| \Delta \tau \right|}
\label{eq:Ctau_2exp_form}
\end{equation}
The autocorrelation function $K$ of the mean square visibility for this impulse-response function then takes the form:
\begin{align}
\frac{N}{C_2 }  \Big\langle K(\left| \Delta \tau \right|) \Big\rangle_S &= \left( \alpha_{1} k_1 (1+k_1 \left| \Delta \tau \right|) -\beta k_2 \right) e^{-k_1 \left| \Delta \tau \right|} +  \left( \alpha_2  k_2 (1+k_2 \left| \Delta \tau \right|) +\beta k_1 \right) e^{-k_2 \left| \Delta \tau \right|} \label{eq:K_2exponential_model} \\
&
\phantom{=}
+
\big( \left( (\alpha_{1} -\beta) k_1 +  (\alpha_{2}+\beta) k_2  \right) \,\delta_0^{\Delta \tau}
 \nonumber
\end{align}
where again $\delta_0^{\Delta\tau}$ is the Kronecker delta function, and:
\begin{align}
 \alpha_1 &= A_1^2  \left( \frac{A_1}{2} + \frac{A_2 k_2}{k_1+k_2}\right)^2 \label{eq:K_2exponential_model_params}\\
 \alpha_2 &= A_2^2  \left( \frac{A_2}{2} + \frac{A_1 k_1}{k_1+k_2}\right)^2 \nonumber \\
 \beta &=  \frac{ A_1 A_2 k_1 k_2 ( 2 A_2 k_2 + A_1 (k_1+k_2)) ( 2 A_1 k_1 + A_2 (k_1+k_2)) }{(k_1-k_2)(k_1+k_2)^3}  \nonumber
\end{align}
Thus, the two exponential scales $1/k_1, 1/k_2$ appear again, in $K$, as do the weights $A_1$, $A_2$.

\section{Anisotropic Scattering in a Thin Screen}\label{sec:anisotropic}

An observer sees an anisotropic distribution of radiation from a screen at distance $D$ from the observer.
The probability of receiving radiation from the screen at position $(\theta_x,\theta_y)$ is:
\begin{align}
P(\theta_x,\theta_y) &= \frac{1}{2\pi \sigma_x \sigma_y} 
\exp\left\{ -\frac{1}{2}\left( \left(\frac{\theta_x}{\sigma_x}\right)^2 +  \left(\frac{\theta_y}{\sigma_y}\right)^2 \right)\right\}\, d\theta_x\, d\theta_y
\label{eq:anis_angle_dist}
\end{align}
We suppose without loss of generality that $\sigma_y > \sigma_x$.
If the source is at infinite distance beyond the screen, the delay $\tau$ along this path (ignoring any contribution from the screen) is:
\begin{align}
c \tau &= D \left(\theta_x^2 + \theta_y^2\right)
\end{align}
where $c$ is the speed of light.
If the source is at distance $R$ beyond the screen, then $D$ is replaced by $RD/(R+D)$ in this and subsequent equations.
We convert the distribution  of angles in Equation\ \ref{eq:anis_angle_dist} to coordinates $(\tau, \phi)$,
where $\phi=\arctan (\theta_y/\theta_x)$.
The resulting distribution of $(\tau,\phi )$ is:
\begin{align}
P(\tau,\phi)\, d\phi\, d\tau &=  \sqrt{1+\alpha^2} \frac{c }{\pi D \sigma_y^2} \exp\left\{ -\frac{c \tau}{2 D \sigma_y^2} \left(1+\alpha^2 \cos^2\phi \right)\right\}\, d\tau\, d\phi
\end{align}
where $\alpha^2=\sigma_y^2/\sigma_x^2 -1$ parametrizes the anisotropy.
We integrate over $\phi$ to find the pulse-broadening function:
\begin{align}
G(\tau) \, d\tau &=\int_0^{2\pi} P(\tau,\phi)\, d\phi\, d\tau 
\end{align}
Thus,
\begin{align}
G(\tau) \, d\tau &= \sqrt{1+\alpha^2}  \exp\left\{ -\frac{(2+\alpha^2)}{2} \frac{\tau}{\tau_2} \right\}\, 
I_0\left( \frac{\alpha^2}{2}  \frac{\tau}{\tau_2}  \right) \frac{d\tau  }{\tau_2}
\end{align}
Here, $I_0$ is the regular modified Bessel function of order 0,
and $c \tau_2=2 D \sigma_y^2$.
Note that for $\alpha\rightarrow 0$, this distribution becomes the familiar exponential distribution, with scale
$\tau_2$, as expected for $\sigma_x=\sigma_y$.  \citet{rickett2009} present a similar expression.

In the case of $\alpha>0$, at small $\tau$ the distribution has the limit:
\begin{align}
\lim_{\tau\rightarrow 0} G(\tau) \, d\tau &\rightarrow \sqrt{1+\alpha^2} \exp\left\{ -\frac{(2+\alpha^2)}{2}\frac{\tau}{\tau_2} \right\}  \frac{d\tau  }{\tau_2}
\end{align}
At large values,
\begin{align}
\lim_{u\rightarrow \infty} I_0(u) \rightarrow \sqrt{\frac{1}{2\pi u}} e^u
\end{align}
so that
\begin{align}
\lim_{\tau\rightarrow \infty} G(\tau) \, d\tau &\rightarrow \sqrt{1+\alpha^2} \sqrt{\frac{1}{2\pi (2+\alpha^2)} \frac{\tau_2}{\tau} }\exp\left\{ -\frac{ \tau}{\tau_2} \right\}  \frac{d\tau  }{\tau_2}
\end{align}
At a particular scale $\tau$, the logarithmic derivative of $G(\tau)$ is $1/\tau_2$, although the coefficient depends weakly on $\tau$.
Thus, $G(\tau)$ exhibits two exponential scales: $\tau_2$ at large $\tau$, and $\tau_1 = \tau_2/(1+\alpha^2/2)$ at small $\tau$.
The relative strength of the scales is about 
\begin{align}
\frac{A_1}{A_2} &= \sqrt{\frac{1+\alpha^2}{2\pi (2+\alpha^2)}  }
\end{align}
near the larger scale $\tau_2$, with the larger scale being the weaker.



\bigskip
\begin{thebibliography}{}
\expandafter\ifx\csname natexlab\endcsname\relax\def\natexlab#1{#1}\fi

\bibitem[{{Andrianov} {et~al.}(2014){Andrianov}, {Guirin}, {Jarov}, {Kostenko},
  {Likhachev}, \& V.}]{RACorrelator}
{Andrianov}, A.~S., {Guirin}, I.~A., {Jarov}, V.~E., {et~al.} 2014, Vestnik
  NPO, 24, 55

\bibitem[{{Bartel} {et~al.}(1985){Bartel}, {Ratner}, {Shapiro}, {Cappallo},
  {Rodgers}, \& {Whitney}}]{bartel1985}
{Bartel}, N., {Ratner}, M.~I., {Shapiro}, I.~I., {et~al.} 1985, \aj, 90, 2532

\bibitem[{{Boldyrev} \& {Gwinn}(2003)}]{boldyrev2003}
{Boldyrev}, S., \& {Gwinn}, C.~R. 2003, Physical Review Letters, 91, 131101

\bibitem[{{Bracewell}(2000)}]{bracewell2000}
{Bracewell}, R.~N. 2000, {The Fourier transform and its applications} (McGraw
  Hill)

\bibitem[{Brisken {et~al.}(2002)Brisken, Benson, Goss, \&
  Thorsett}]{brisken2002}
Brisken, W.~F., Benson, J.~M., Goss, W.~M., \& Thorsett, S.~E. 2002, \apj, 571,
  906

\bibitem[{{Brisken} {et~al.}(2010){Brisken}, {Macquart}, {Gao}, {Rickett},
  {Coles}, {Deller}, {Tingay}, \& {West}}]{brisken2010}
{Brisken}, W.~F., {Macquart}, J.-P., {Gao}, J.~J., {et~al.} 2010, \apj, 708,
  232

\bibitem[{{Britton} {et~al.}(1998){Britton}, {Gwinn}, \& {Ojeda}}]{britton1998}
{Britton}, M.~C., {Gwinn}, C.~R., \& {Ojeda}, M.~J. 1998, \apjl, 501, L101

\bibitem[{{Desai} {et~al.}(1992){Desai}, {Gwinn}, {Reynolds}, {King},
  {Jauncey}, {Flanagan}, {Nicolson}, {Preston}, \& {Jones}}]{desai1992}
{Desai}, K.~M., {Gwinn}, C.~R., {Reynolds}, J., {et~al.} 1992, \apj, 393, L75

\bibitem[{{Edwards} {et~al.}(2006){Edwards}, {Hobbs}, \& {Manchester}}]{TEMPO}
{Edwards}, R.~T., {Hobbs}, G.~B., \& {Manchester}, R.~N. 2006, \mnras, 372,
  1549

\bibitem[{{Goodman} \& {Narayan}(1989)}]{goodman1989}
{Goodman}, J., \& {Narayan}, R. 1989, \mnras, 238, 995

\bibitem[{Gwinn}(2001)]{gwinn2001}{{Gwinn}, C.~R.}, 2001, \apj, 554,1197


\bibitem[{{Gwinn} {et~al.}(2006){Gwinn}, {Hirano}, \& {Boldyrev}}]{gwinn2006}
{Gwinn}, C.~R., {Hirano}, C., \& {Boldyrev}, S. 2006, \aap, 453, 595

\bibitem[{{Gwinn} \& {Johnson}(2011)}]{GwinnJohnson2011}
{Gwinn}, C.~R., \& {Johnson}, M.~D. 2011, \apj, 733, 51

\bibitem[{{Gwinn} {et~al.}(2011){Gwinn}, {Johnson}, {Smirnova}, \&
  {Stinebring}}]{0834noise2011}
{Gwinn}, C.~R., {Johnson}, M.~D., {Smirnova}, T.~V., \& {Stinebring}, D.~R.
  2011, \apj, 733, 52

\bibitem[{{Gwinn} {et~al.}(2014){Gwinn}, {Kovalev}, {Johnson}, \&
  {Soglasnov}}]{gwinn2014}
{Gwinn}, C.~R., {Kovalev}, Y.~Y., {Johnson}, M.~D., \& {Soglasnov}, V.~A. 2014,
  \apjl, 794, L14

\bibitem[{{Gwinn} {et~al.}(2012){Gwinn}, {Johnson}, {Reynolds}, {Jauncey},
  {Tzioumis}, {Dougherty}, {Carlson}, {Del Rizzo}, {Hirabayashi}, {Kobayashi},
  {Murata}, {Edwards}, {Quick}, {Flanagan}, \& {McCulloch}}]{velanoise2012}
{Gwinn}, C.~R., {Johnson}, M.~D., {Reynolds}, J.~E., {et~al.} 2012, \apj, 758,
  6

\bibitem[{{Hagfors}(1976)}]{Hagfors1976}
{Hagfors}, T. 1976, Methods of Experimental Physics, 12, 119

\bibitem[{{Hewish}(1980)}]{hewish1980}
{Hewish}, A. 1980, \mnras, 192, 799

\bibitem[Johnson et al.(2016)]{3C273_RA}
Johnson, M.~D., Kovalev, Y.~Y., Gwinn, C.~R., et al.\ 2016, \apjl, 820, L10

\bibitem[{{Kardashev} {et~al.}(2013){Kardashev}, {Khartov}, {Abramov},
  {Avdeev}, {Alakoz}, {Aleksandrov}, {Ananthakrishnan}, {Andreyanov},
  {Andrianov}, {Antonov}, {Artyukhov}, {Arkhipov}, {Baan}, {Babakin},
  {Babyshkin}, {Bartel'}, {Belousov}, {Belyaev}, {Berulis}, {Burke},
  {Biryukov}, {Bubnov}, {Burgin}, {Busca}, {Bykadorov}, {Bychkova},
  {Vasil'kov}, {Wellington}, {Vinogradov}, {Wietfeldt}, {Voitsik},
  {Gvamichava}, {Girin}, {Gurvits}, {Dagkesamanskii}, {D'Addario},
  {Giovannini}, {Jauncey}, {Dewdney}, {D'yakov}, {Zharov}, {Zhuravlev},
  {Zaslavskii}, {Zakhvatkin}, {Zinov'ev}, {Ilinen}, {Ipatov}, {Kanevskii},
  {Knorin}, {Casse}, {Kellermann}, {Kovalev}, {Kovalev}, {Kovalenko}, {Kogan},
  {Komaev}, {Konovalenko}, {Kopelyanskii}, {Korneev}, {Kostenko}, {Kotik},
  {Kreisman}, {Kukushkin}, {Kulishenko}, {Cooper}, {Kut'kin}, {Cannon},
  {Larionov}, {Lisakov}, {Litvinenko}, {Likhachev}, {Likhacheva}, {Lobanov},
  {Logvinenko}, {Langston}, {McCracken}, {Medvedev}, {Melekhin}, {Menderov},
  {Murphy}, {Mizyakina}, {Mozgovoi}, {Nikolaev}, {Novikov}, {Novikov},
  {Oreshko}, {Pavlenko}, {Pashchenko}, {Ponomarev}, {Popov}, {Pravin-Kumar},
  {Preston}, {Pyshnov}, {Rakhimov}, {Rozhkov}, {Romney}, {Rocha}, {Rudakov},
  {R{\"a}is{\"a}nen}, {Sazankov}, {Sakharov}, {Semenov}, {Serebrennikov},
  {Schilizzi}, {Skulachev}, {Slysh}, {Smirnov}, {Smith}, {Soglasnov},
  {Sokolovskii}, {Sondaar}, {Stepan'yants}, {Turygin}, {Turygin}, {Tuchin},
  {Urpo}, {Fedorchuk}, {Finkel'shtein}, {Fomalont}, {Fejes}, {Fomina},
  {Khapin}, {Tsarevskii}, {Zensus}, {Chuprikov}, {Shatskaya}, {Shapirovskaya},
  {Sheikhet}, {Shirshakov}, {Schmidt}, {Shnyreva}, {Shpilevskii}, {Ekers}, \&
  {Yakimov}}]{kardashev2013}
{Kardashev}, N.~S., {Khartov}, V.~V., {Abramov}, V.~V., {et~al.} 2013,
  Astronomy Reports, 57, 153

\bibitem[{{Kondratiev} {et\ al.}(2007){Kondratiev}, {Popov}, {Soglasnov},
  {Kovalev}, {Bartel}, {Cannon}, \& {Novikov}}]{kondratiev2007}
{Kondratiev}, V.~I., {Popov}, M.~V., {Soglasnov}, V.~A., {et~al.} 2007,
  Astronomical and Astrophysical Transactions, 26, 585

\bibitem[{{Lorimer} {et~al.}(1995){Lorimer}, {Yates}, {Lyne}, \&
  {Gould}}]{Lorimer1995}
{Lorimer}, D.~R., {Yates}, J.~A., {Lyne}, A.~G., \& {Gould}, D.~M. 1995,
  \mnras, 273, 411
  
  \bibitem[{{Narayan}(1992)}]{narayan1992}{Narayan}, R. \& {Goodman}, J. 1989, {Philosophical Transactions of the Royal Society of London Series A}, 341, 151

\bibitem[{{Narayan} \& {Goodman}(1989)}]{narayan1989}
{Narayan}, R., \& {Goodman}, J. 1989, \mnras, 238, 963

\bibitem[{{Pen} \& {Levin}(2014)}]{PenLevin2014}
{Pen}, U.-L., \& {Levin}, Y. 2014, \mnras, 442, 3338

\bibitem[{{Popov} \& {Soglasnov}(1984)}]{popov1984}
{Popov}, M.~V., \& {Soglasnov}, V.~A. 1984, \sovast, 28, 424

\bibitem[{Prokhorov {et~al.}(1975)Prokhorov, Bunkin, Gochelashvily, \&
  Shishov}]{prokhorov1975}
Prokhorov, A., Bunkin, V.~F., Gochelashvily, K.~S., \& Shishov, V.~I. 1975,
  Proc. IEEE, 63, 790

\bibitem[{{Rickett}(1975)}]{rickett1975}
{Rickett}, B.~J. 1975, \apj, 197, 185

\bibitem[{{Rickett}(1977)}]{rickett1977}
---. 1977, \araa, 15, 479

\bibitem[{{Rickett et\ al.}(2009)}]{rickett2009}
{Rickett}, B., {Johnston}, S., {Tomlinson}, T., \& {Reynolds}, J. 2009, \mnras, 395, 1391


\bibitem[{{Semenkov} {et~al.}(2004){Semenkov}, {Soglasnov}, \&
  {Popov}}]{semenkov2004}
{Semenkov}, K.~V., {Soglasnov}, V.~A., \& {Popov}, M.~V. 2004, Astronomy
  Reports, 48, 457

\bibitem[{{Shishov} {et~al.}(2003){Shishov}, {Smirnova}, {Sieber}, {Malofeev},
  {Potapov}, {Stinebring}, {Kramer}, {Jessner}, \& {Wielebinski}}]{shishov2003}
{Shishov}, V.~I., {Smirnova}, T.~V., {Sieber}, W., {et~al.} 2003, \aap, 404,
  557

\bibitem[{{Smirnova} {et~al.}(2014){Smirnova}, {Shishov}, {Popov}, {Gwinn},
  {Anderson}, {Andrianov}, {Bartel}, {Deller}, {Johnson}, {Joshi}, {Kardashev},
  {Karuppusamy}, {Kovalev}, {Kramer}, {Soglasnov}, {Zensus}, \&
  {Zhuravlev}}]{smirnova2014}
{Smirnova}, T.~V., {Shishov}, V.~I., {Popov}, M.~V., {et~al.} 2014, \apj, 786,
  115

\bibitem[{Stinebring {et~al.}(2001)Stinebring, McLaughlin, Cordes, Becker,
  Goodman, Kramer, Sheckard, \& Smith}]{stinebring2001}
Stinebring, D.~R., McLaughlin, M.~A., Cordes, J.~M., {et~al.} 2001, \apj, 549,
  L97

\bibitem[{{Thompson} {et~al.}(2007){Thompson}, {Moran}, \& {Swenson}}]{TMS2007}
{Thompson}, A.~R., {Moran}, J.~M., \& {Swenson}, G.~W. 2007, {Interferometry
  and Synthesis in Radio Astronomy, John Wiley \& Sons, 2007.} (Wiley)

\bibitem[{{Weisstein}(2014)}]{FTConvention}
{Weisstein}, E.~W. 2014, Fourier Transform,
  \url{http://mathworld.wolfram.com/FourierTransform.html}, accessed:
  2015-09-02

\bibitem[{{Yangalov} {et~al.}(2001){Yangalov}, {Popov}, {Soglasnov},
  {Semenkov}, {Hirabayashi}, {Liu}, \& {Wang}}]{yangalov2001}
{Yangalov}, A.~K., {Popov}, M.~V., {Soglasnov}, V.~A., {et~al.} 2001, \apss,
  278, 39

\end{thebibliography}

\end{document}